\documentclass[pra,
a4paper,
twocolumn,
superscriptaddress]{revtex4-1}

\usepackage{amssymb}
\usepackage{amsmath}
\usepackage{amsfonts}
\usepackage{braket}
\usepackage{graphicx}
\usepackage{bm}
\usepackage{color}
\usepackage{multirow}
\usepackage{natbib}
\usepackage{hyperref}
\usepackage{verbatim}
\usepackage{dcolumn}
\usepackage{ulem}
\usepackage{float}

\def \be {\begin{equation}}
\def \ee {\end{equation}}

\newcommand{\aleq}[1]{
\begin{equation}
    \begin{aligned}
    #1
    \end{aligned}
\end{equation}
}

\begin{document}

\title{Shot noise in Aharonov-Bohm interferometers: \\ 
Comparison of helical and conventional setups}








\author{R.A. Niyazov}
\affiliation{Ioffe Institute, St. Petersburg 194021, Russia}
\affiliation{NRC ``Kurchatov Institute'', Petersburg Nuclear Physics Institute, Gatchina 188300, Russia}
\email{niyazov\_ra@pnpi.nrcki.ru} 
\author{I.V. Krainov}
\affiliation{Ioffe Institute, St. Petersburg 194021, Russia}
\author{D.N. Aristov}
\affiliation{Ioffe Institute, St. Petersburg 194021, Russia}
\affiliation{NRC ``Kurchatov Institute'', Petersburg Nuclear Physics Institute, Gatchina 188300, Russia}
\affiliation{Department of Physics, St. Petersburg State University, St. Petersburg 199034, Russia}
\author{V.Y. Kachorovskii}
\affiliation{Ioffe Institute, St. Petersburg 194021, Russia}

\begin{abstract}
We study tunneling transport through quantum Aharonov-Bohm (AB) interferometers and demonstrate that interference effects strongly modify shot noise of the current. We discuss in detail two simplest setups: conventional single-channel spinless interferometer and interferometer formed by helical edge states of two-dimensional topological insulator. We demonstrate that both in the conventional and the helical case the interference dramatically changes the Fano factor and its magnetic field dependence. For weak tunneling coupling, the Fano factor of both setups exhibits a periodic series of sharp AB peaks depending on the magnetic flux piercing the system. Our key finding is that the Fano factor in the helical interferometer provides information about the presence of backscattering defects violating topological protection. In particular, the amplitude of AB peaks in the helical setup is proportional to the strength of the defect in contrast to conventional setup, where peaks have finite amplitude even in the ballistic case.
\end{abstract}


\maketitle

\section{Introduction}

Quantum interferometry  is a rapidly growing area of electronics with  a huge  potential for practical use, in particular, in the booming  fields of quantum computing and quantum cryptography.  
One of the beautiful manifestation of quantum interference is the Aharonov-Bohm effect \cite{Aharonov1959b, Aharonov1963}---the  sensitivity of the phase of electron wave function to magnetic flux. Quantum interferometers based on this effect are used for high-precision measurements of magnetic fields, for example, by using SQUID circuits. 

The   simplest  setup to study interference and particularly the Aharonov-Bohm effect  is a ring-shaped  interferometer with two single-channel arms placed in the perpendicular magnetic field. Due to
the interference of electron trajectories winding
around such a setup, the observables, like transmission   coefficient and 
the noise intensity, show Aharonov-Bohm oscillations with a certain periods $\Delta \phi$     
as a function of  dimensionless magnetic flux $\phi=\Phi/\Phi_0$
threading the device, where $\Phi_0 = hc/e$ is the flux quantum and $\Phi=B S,$ where $B$ is homogeneous magnetic field and $S$ is  the area of the region encompassed  by one-dimensional channel. The key statement of our paper is that noise in such a simple system is still  not well understood  and depends on specific realizations of the single-channel arms.

\begin{figure}
    \centering
    \includegraphics[width=0.8\linewidth]{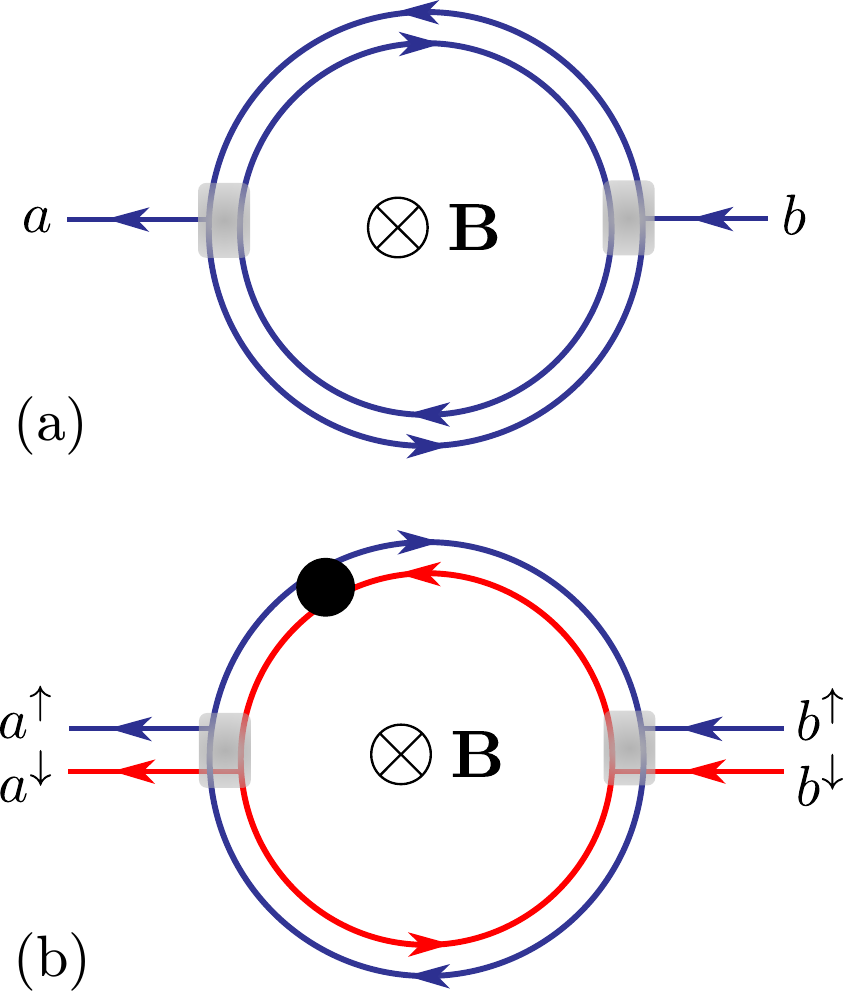}
\caption{Aharonov-Bohm interferometers based on conventional spinless single-channel quantum wire (a) and on helical edge channels of 2D topological insulator (b). The  leads  (shown by gray color) are modeled by conventional single-channel [spinless (a) and spinful (b)] wires. The black dot in panel (b) represents  a    backscattering     impurity.  {Gray regions represent contacts described by $3\times 3$ and $4\times 4$ $S-$matrix} for CI and HI, respectively [see Eqs.~\eqref{eq:SCI}, \eqref{S-matrix} and Figs.~\ref{Fig-conv-lead}, \ref{fig:lead}].  Homogeneous magnetic field, $B,$  is perpendicular to the picture plane.  Corresponding  magnetic flux,  $\Phi=B S,$  is proportional to  the area, $S,$ of the region encompassed  by one-dimensional channels}.   
\label{fig:setup}
\end{figure}

Here, we will investigate   and compare in details   two types of setups:  (i) a spinless single-channel  interferometer based on conventional materials,  i.e.  a single-channel spinless  quantum wire, folded in a  ring-like shape and having  two tunnel or metallic contacts attached (see Fig.~\ref{fig:setup}a); (ii) an interferometer based on helical edge states. Below we refer to these systems as a conventional interferometer (CI) and a helical interferometer (HI), respectively.
The latter interferometer  can be implemented in two-dimensional topological insulators ~\cite{Bernevig2013}, which are insulators in the bulk, but  have helical one-dimensional (1D) edge states  conducting the current  without dissipation \cite{Bernevig2013,Hasan2010,Qi2011}. These edge states are formed by electrons with opposite spins traveling in opposite directions, so that  backscattering by non-magnetic impurities is prohibited. 
If one attaches two contacts to the edge  and apply the gate voltage to shift  the Fermi level into the band gap, then the conductance  of such a device is fully determined by the  edge states (see Fig.~\ref{fig:setup}b). Since the sample boundary can be bypassed  in two directions, possibly with several windings,   such a system represents  an interferometer that encodes  information about properties  of  the helical edge states (HES).

Both setups, CI and HI, have two electron waves propagating clockwise and counterclockwise in 1D  channel. However, the interference phenomena in these systems 
 manifest themselves
  in a rather different way. 
 
A  spinless CI based on conventional materials (Fig.~\ref{fig:setup}a) shows interference effects even  in the simplest  ballistic case, when both arms of interferometer do not contain any  impurities.  By contrast,      
the  waves, propagating in opposite directions in ballistic HI, have also opposite spin projections and cannot interfere provided that contacts to the HES  are non-magnetic (note that CI differs from Rashba-interferometers, where ballistic interference is possible due to the Aharonov-Casher effect \cite{Aronov1993,Koga2004,Nitta2007,Shmakov2012}, see also  Ref.~\cite{Manchon2015}  for review   and references therein). 
Interference effects in HI  appear in    presence of at least one  backscattering center (see Fig.~\ref{fig:setup}b).  It means that the study of interference effects can help to solve the well-known problem of the lack of topological protection in real samples (see  recent experiment \cite{Olshanetsky2023a} and discussion  of the experimental situation therein).  Specifically,  observation of  interference phenomena in the HI can give information about mechanisms of possible violation of the topological protection.

For both setups we will  investigate the manifestations of  quantum interference in the 
shot noise of the current. Such noise is a consequence of the discreteness of the  electron charge. 
Importantly, its measurement provides information not available from conductance measurements: namely, the charge and statistics of current carriers, and the internal energy scales of the system~\cite{Jong1997,Blanter2000}.  Our primary goal is to calculate    the Fano factor, $
\mathcal F$, which is the ratio of the shot noise and the 
so-called Schottky noise value.
This ratio is often used to characterize different transport regimes.    
For example,  the values $\mathcal F = 1/3~$  and  $\mathcal F=1/2~$  for a diffusive conductor and  tunneling  transmission through  a quantum  level, respectively,  are universal and independent of the system details~\cite{Jong1997,Blanter2000}.  As we demonstrate below,  the Fano factor also is very sensitive to the interference effects which manifest itself in different way in CI and HI.

It is worth  noting that the shot noise in HES has already been discussed  for the case of infinite edge  \cite{Lezmy2012, DelMaestro2013, Tikhonov2015, Mani2017, Vaeyrynen2017,Piatrusha2018,Kurilovich2019, Kurilovich2019a,Pashinsky2020,Hsu2021,Probst2022,Mishra2023}, i.e. without taking into  account the  interference effects. 
In particular, a great attention 
 was paid
 to infinite HES with a dynamic magnetic impurity \cite{Vaeyrynen2017,Kurilovich2019,Kurilovich2019a,Pashinsky2020,Hsu2021,Probst2022}),  whose magnetic moment   changes direction after each scattering event. 
Discussing HI in this paper, we focus on the opposite case of static backscattering  defect, which  is not necessarily magnetic (in particular,  backscattering mechanisms involving charged puddles in the bulk of topological insulator are actively discussed \cite{Vaeyrynen2013}). We will describe such defect by $S$ matrix of general type.    The experimental measurement of the Fano factor of HI at zero magnetic field for the edge states of 2D TI gives the value $0.1 <\mathcal F<0.3$ ~\cite{Tikhonov2015,Piatrusha2018}. The upper value, 0.3, is close to the value  $1/3$  for diffusive conductor.
A similar result was obtained  in the model with a large number of ``islands'',  allowing spin relaxation and  tunnel-connected to the HES   \cite{Aseev2016}. The effect  of  islands of different types   on HES is currently being actively debated (see also recent works \cite{Olshanetsky2023a,Krainov2025} and references therein).

Since the  conductance and shot noise of interferometers of various types  were  also studied in a great number of works,
we will start with a short description of the state of the art.  Let us  first   recall the key results known for the conductance of the Aharonov-Bohm interferometer.   
At weak tunnel coupling and at low temperatures, the conductance $G(\phi)$ exhibits sharp resonances at certain values of magnetic flux, arising when quantum levels of the system are crossed by the Fermi level  \cite{Buettiker1984,Buettiker1985}. 
  
As was understood  later \cite{Jagla1993}, 
the case of relatively high temperatures, $T$, is much less trivial. 
One might suggest that interference is suppressed when $T$ becomes larger than the level spacing, $\Delta.$
However, as was first predicted in  Ref.~\cite{Jagla1993} for CI, the  conductance  shows interesting interference-induced  behavior in this case.  Specifically,
as it was  shown theoretically, for  
both   CI \cite{Jagla1993,Dmitriev2010,Shmakov2012,Shmakov2013, Dmitriev2015b,Dmitriev2017} and    HI \cite{Niyazov2018, Niyazov2020,Niyazov2021,Niyazov2021a},   the interference effects  lead to  resonance behavior of the conductance $G$  even in the case 
\be T\gg \Delta= 2\pi v_{ \rm F}/L, 
\label{cond1}
\ee 
where $L= L_1+L_2,$ with $L_{1,2}$  the lengths of the interferometer arms, and $v_{\rm F}$ is the Fermi velocity. However, instead of resonances, $G(\phi)$ exhibits narrow antiresonances under the condition  \eqref{cond1}. Physically, anti-resonances arise due to trajectories that interfere destructively at any energy $\varepsilon$ and, accordingly, are insensitive to energy averaging (see  discussions in \cite{Dmitriev2010} and \cite{Niyazov2018} regarding  CI and HI, respectively).

The conductance in CI in the temperature range \eqref{cond1} has been studied in detail, in particular, considering the effects of the electron-electron interaction \cite{Dmitriev2010},
disorder \cite{Shmakov2013}, and 
spin-orbit interaction \cite{Shmakov2012}. Recently,  some of these results were generalized to HI systems: to single ring-shaped HI \cite{Niyazov2018, Niyazov2020,Niyazov2021,Niyazov2021a} with a static magnetic defect and, partially, to arrays of coupled helical rings \cite{Niyazov2023}. (See also discussion for other types of interferometers \cite{Feldman2007, Wang2010, Yang2015}.)

Surprisingly,  non-trivial interference-induced characteristics of noise emerging under condition  \eqref{cond1} were not discussed previously even for the simplest possible model of a single-channel CI, in view of the huge number of publications devoted to CI. To the best of our knowledge, the influence of the interference effects on the noise has previously been discussed only in the low-temperature regime, $T \ll \Delta$, see review in  Ref.~\cite{Kobayashi2021}  for  CI and  Refs.~ \cite{Edge2013,Dolcini2015} for HI.

On the other hand there is a clear need of the corresponding theory both for CI and HI, which would  allow direct comparison with experiment. Indeed, nanoscale rings made from conventional materials with a few or single conducting channels have been already fabricated \cite{Shea2000,Piazza2000,Fuhrer2001,Keyser2003,Zou2007}. Also, the  experiment aimed to study  HI was recently reported \cite{Munyan2023}, although it focused on  the  low-temperature case. 
Note that in the latter experiment a quantum point contacts to the HES have been realized, thus allowing experimental verification of interference effects in conductance and noise of tunneling HI, especially having in mind that the condition \eqref{cond1} does not require very strict restrictions on temperature.
Indeed, for  typical $v_{\rm F}$ of  order of $10^7~$cm/s and typical system sizes ($> 1$ micron), the value of $\Delta$ does not exceed several degrees Kelvin. This means that the interference effects  can be studied at relatively high temperatures, which are relevant for various applications.   
A comment about  magnetic-field-induced  phenomena  is in order to do. Actually, the magnetic field causes two effects: (i) the AB phase appears and  (ii) the topological protection is destroyed. 
Simple estimates show that the flux quantum,  $\Phi \sim \Phi_0 = hc/e,$ is achieved for sample with micron   size in fields  $B  \sim  3$ Oe, well below the expected magnitude of the fields destroying the edge states \cite{Du2015,Zhang2014,Hu2016a}.

Recently we discussed shot noise in HI \cite{Niyazov2024}. 
We considered  HI with identical contacts  and  calculated the noise power in the regime \eqref{cond1} paying attention to the interference effects.   We found  the Fano factor strongly depending on the tunneling coupling and strength of backscattering defect. 

Here, we further develop the results obtained in Ref.~\cite{Niyazov2024}. We present a detailed study of the sharp  Aharonov-Bohm peaks in the  Fano factor of the noise.    
We  also  demonstrate that the simultaneous measurement of the conductance and the Fano factor of HI allows one to directly determine the strength of the backscattering defect 
even in the absence of detailed information about scattering amplitudes characterizing the contacts.
 Having also in mind the recent experiment \cite{Munyan2023}, we generalize here the calculation of the Fano factor of HI for experimentally relevant case of non-equivalent contacts.       

In Ref.~\cite{Niyazov2024}, we also briefly discussed the case of CI, announcing without derivation the formula for the Fano factor of the CI with symmetric arms and  identical contacts. In this paper we provide detailed calculations, and generalize our analysis for the experimentally  relevant case   of  asymmetric arms and  non-identical contacts \cite{Shea2000,Piazza2000,Fuhrer2001,Keyser2003,Zou2007}.  We study in detail the dependence of the noise of a CI on the magnetic field.  It turns out that the interference effects are  much more  pronounced  for CI and, consequently,  the influence of the magnetic field on the Fano factor is much stronger as compared to HI.  The reason is that the  interference  comes into play for CI already in the ballistic case, when impurities in both arms of CI are absent, while the Aharonov-Bohm peaks in  HI are proportional to the backscattering probability.

 We will discuss  ballistic CI without impurities and defects   and also   HI with  a single static defect (which can be magnetic or non-magnetic) placed into one of the arm of the interferometer.
The goal of the work is to find the dependence of Fano factor, $\mathcal F$,       
on tunneling coupling,  on  the magnetic flux,  and, in the case of HI, on the strength of backscattering defect. 
We will  find  general analytical expressions for  
$\mathcal F^{\rm CI}$,   $\mathcal F^{\rm HI}$ and analyze simple limiting cases assuming that  inequality \eqref{cond1} is satisfied.

\section{Model}
We   calculate the  current  shot noise in the Aharonov-Bohm (AB) interferometers of two types, CI  and   HI,   depicted in  Fig.~\ref{fig:setup}a and  Fig.~\ref{fig:setup}b, respectively. They are based on spinless single-channel quantum wire and helical edge states.
We model leads by conventional single-channel wires  (spinless or spinful for the conventional  and   helical case, respectively).   Although this model of the leads is very simplified, it is commonly used in  quantum interferometry, starting from the work  \cite{Buettiker1984}, since,   as we demonstrate below, it allows  describing the transition from the metallic leads  to the tunnel ones. In particular, this model qualitatively describes quantum point contacts to the  interferometer.  
  
\subsection{Conventional AB interferometer}
  
\begin{figure}
	\centering
 \includegraphics[width=0.35\linewidth]{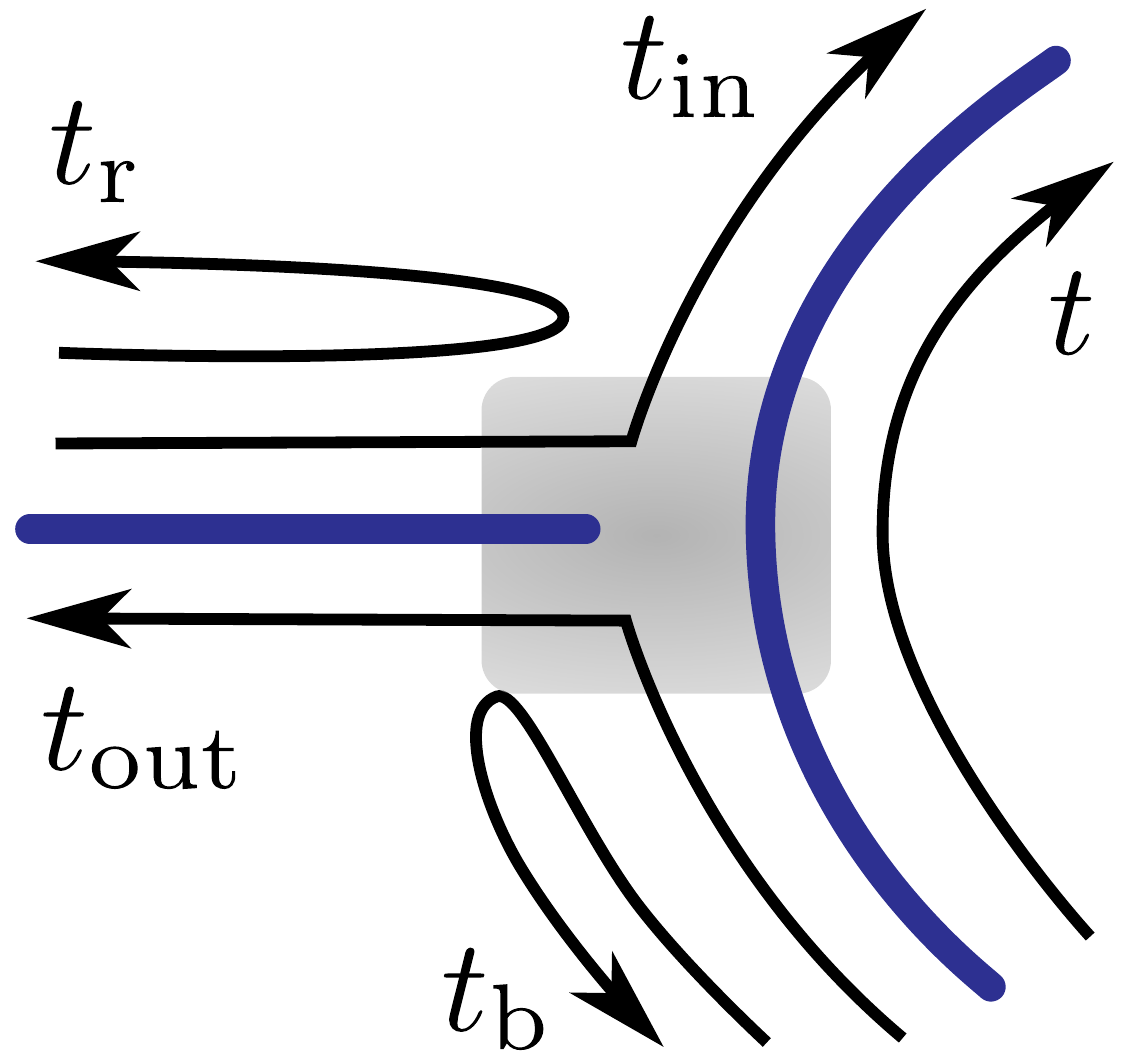}
	\caption{Scattering amplitudes entering scattering matrix Eq.~\eqref{eq:SCI}  for CI.  The lead is modeled by spinless 1D wire.  For tunneling contact, $|t_{\rm r}| \approx 1,$ while the case of metallic contact  corresponds to $t_{\rm r}=0.$ Contact region with area $S_0$ is shown by gray color.  We assume that $B S_0 \ll \Phi_0,$ so that the magnetic field does not affect S-matrix.}
	\label{Fig-conv-lead}
\end{figure}

The unitarity of scattering matrix of the contact results in backscattering on the contacts  in case of CI \cite{Dmitriev2010}.
Here, we use this scattering matrix with real-valued elements: 
\aleq{
\hat S^{\mathrm{CI}}=&
\begin{pmatrix}
t_{\mathrm{r}} & t_{\mathrm{out}} & t_{\mathrm{out}} \\
t_{\mathrm{in}} & t_{\mathrm{b}} & t \\
t_{\mathrm{in}} & t & t_{\mathrm{b}}\\
\end{pmatrix} 
\, , \\
t=& \frac{1}{1+\gamma}\,, \quad t_{\mathrm{b}} = -\frac{\gamma}{1+\gamma} \,,\\
t_{\mathrm{in}}=&t_{\mathrm{out}}=\frac{\sqrt{2 \gamma}}{1+\gamma}\,, \quad t_{\mathrm{r}}= - \frac{1-\gamma}{1+\gamma}\,.
\label{eq:SCI}
}
More general case of $\hat S^{\mathrm{CI}}$ is discussed in  Refs.~\cite{Aristov2011a} and \cite{Aristov2010}.
The scattering  amplitudes entering Eq.~\eqref{eq:SCI} obey $ 2 t_{\rm in}^2 + t_{\rm r}^2= t^2 + t_{\rm b}^2 + t_{\rm out}^2=1.$
The meaning of  these amplitudes is illustrated in Fig.~\ref{Fig-conv-lead}: 
 $t_{\rm in} $ is the amplitude  to  enter    the interferometer  into one of the counter-propagating channels, $t_{\rm out}$ is the amplitude to exit the interferometer,    $t$ and $t_{\rm b}$ are respectively the   amplitudes of forward and backward scattering on the contact  inside the interferometer,  and $t_{\rm r }$ is the amplitude of  backward scattering  for electrons  coming from     the lead.      All scattering  amplitudes are  parameterized by a single parameter     $\gamma \in (0, \infty) $. 

\begin{figure}
    \centering
     \includegraphics[width=1\linewidth]{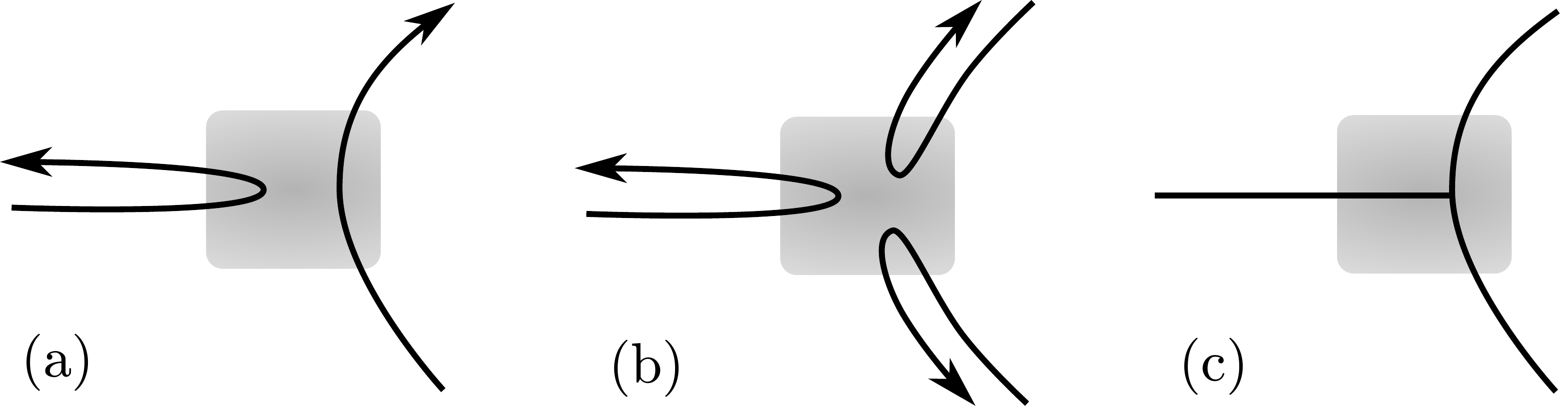}
\caption{Schematic illustration of different types of leads (contact area is shown by gray color): panels (a) and  (b)  shows two different types of tunneling coupling, corresponding to $\gamma=0$ and $\gamma=\infty,$  respectively;  panel (c) illustrates ``metallic''  contact, $\gamma \approx 1.$  Arrows in  panels (a) and (b) show processes with non-zero amplitudes (compare with Fig.~\ref{Fig-conv-lead}). Panel (c) corresponds to a ``metallic contact'', so that all processes   in Fig.~\ref{Fig-conv-lead} are allowed. Scattering matrices, describing cases (a), (b), and (c)  are given by Eq.~\eqref{eq:3cases}. }
\label{fig:contact_conv}
\end{figure}

The meaning of the  parameter  $\gamma $ is further illustrated in Fig.\  ~\ref{fig:contact_conv}, see  also discussion  in Refs.~\cite{Dmitriev2010,Aristov2010,Aristov2017}. The case $\gamma \ll 1 $ corresponds to tunneling contact, where almost closed   interferometer is weakly coupled to the leads,   with $\gamma $ being a tunneling transparency (see Fig.~\ref{fig:contact_conv}a).   Interestingly, the case $\gamma \to \infty$ also corresponds to weak tunneling coupling.  However, this case is strongly different from the case  $\gamma \to 0,$ as illustrated in   Fig.~\ref{fig:contact_conv}b.  Metallic contact is modeled by $\gamma \approx 1$ (see Fig.~\ref{fig:contact_conv}c). The scattering matrices describing   these three cases are given by  (see  Eq.~\eqref{eq:SCI}):
\aleq{\label{eq:3cases}
\hat S^{\mathrm{CI}}(\gamma=0)=&
\begin{pmatrix}
-1 & 0 & 0\\
0& 0 & 1 \\
0 & 1 & 0\\
\end{pmatrix} 
\, , \, 
\hat S^{\mathrm{CI}}(\gamma=\infty)=&
\begin{pmatrix}
1 & 0 & 0 \\
0 & -1 & 0 \\
0 & 0 & -1\\
\end{pmatrix} 
\, , \\
\hat S^{\mathrm{CI}}(\gamma=1)=&
\begin{pmatrix}
0 & \sqrt{2} & \sqrt{2} \\
\sqrt{2} & -1& 1 \\
\sqrt{2} & 1 & -1\\
\end{pmatrix} /2
\, . 
}
It is worth noting that $t_b \neq 0$ for any non-zero $\gamma.$ 
{Below we show that two configurations,  Fig.~\ref{fig:contact_conv}a and Fig.~\ref{fig:contact_conv}b,  may correspond to the same conductance but to different noise. } 

 \subsection{Helical AB interferometer}
  Next we discuss  model of   HI. 
We consider 2D TI, assuming that the Fermi level is located in the bulk gap, so that the transport between two contacts connected to the system is completely determined by the HES.
 \begin{figure}
    \centering
     \includegraphics[width=0.9\linewidth]{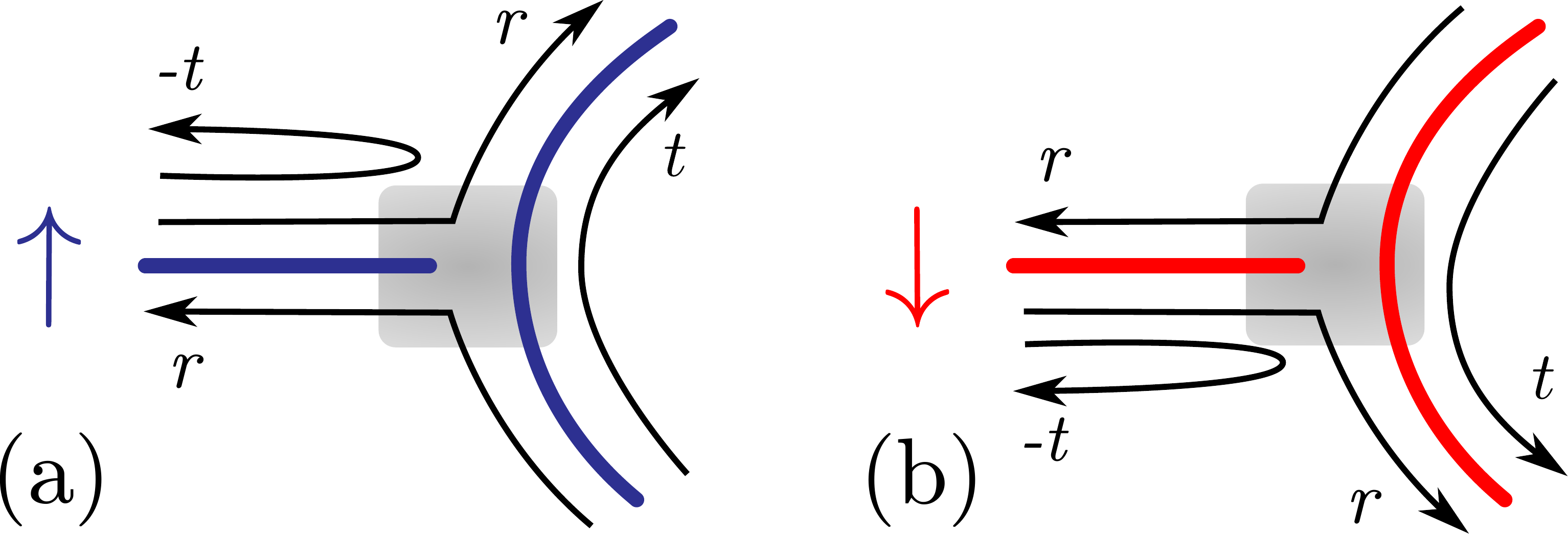}
\caption{The scattering amplitudes entering scattering matrix of contacts, Eq.~\eqref{S-matrix}, which  are modeled as a single-channel spinful wire. It is assumed that there is no spin flip at the contact and the two spin polarizations at the contact (shown by red and blue color) are completely separated. The $t\to 1$ case corresponds to a tunnel contact, and the $t\to 0$ case models a metal contact. 
Contact region with area $S_0$ is shown by gray color.  We assume that $B S_0 \ll \Phi_0,$ so that the field does not affect S-matrix.
}
\label{fig:lead}
\end{figure} 
The simplest  scattering matrix of a non-magnetic  lead has the form
\begin{equation}
\hat{S}^\mathrm{HI}
=  
\left( \begin{array}{cccc}
-t & r &0 &0 \\
r & t &0 &0\\
0&0&-t & r \\
0&0&r & t \\
\end{array} \right) \,,\quad t^2+r^2=1,
\label{S-matrix}
\end{equation}
where two identical blocks are responsible for two spins, and the basis is chosen in accordance with the spin polarization of the helical states at the point of contact (red and blue in Fig. \ref{fig:lead}). It is convenient to parameterize amplitudes $t$ and $r$ by  a parameter $\lambda$:
\be 
t=e^{-\lambda},\quad r=\sqrt{1-e^{-2\lambda}},\quad  0<\lambda<\infty 
\label{helical_t_r_via_lambda}.
\ee
  
Due to the topological protection,     backscattering appears in  the HI    only in  presence of  a  defect that violate time-reversal symmetry, in particular, in  presence of   magnetic defects or in  case of magnetization of the leads.   
We  assume in this paper  that the leads are non-magnetic, but there is a static backscattering  defect in one of the arms of the interferometer. 
As we already mentioned, much attention was drawn recently to the infinite HES with a dynamic magnetic impurity  \cite{Vaeyrynen2017,Kurilovich2019,Kurilovich2019a,Pashinsky2020,Hsu2021,Probst2022}),  whose magnetic moment  changes direction after each scattering event.  For the isotropic exchange interaction between the impurity and the HES,  the impurity's magnetic moment  relaxes within the so-called Korringa time towards the local direction of the electron spin in the HES and, as a consequence, the interaction between the impurity and the HES becomes completely ineffective (see discussion in \cite{Vaeyrynen2017}). Accordingly, the problem of the noise intensity at zero frequency in the HES with a single impurity makes sense only in the presence of an anisotropic exchange interaction \cite{Kurilovich2019} (or an external magnetic field acting on a dynamic impurity \cite{Probst2022}), while in the  limit of isotropic exchange interaction 
 Fano factor  is singular  \cite{Kurilovich2019}, i.e.  the result for the $\mathcal F$ depends on the order in which the constants responsible for the anisotropy tend to zero (see also discussion in \cite{Nagaev2018})

At the same time, the relaxation of the impurity's magnetic moment  in reality stems not only from the interaction with the HES, but also from the environment of the impurity, which should provide a non-singular response for the $\mathcal F$ even for the isotropic exchange interaction. Therefore, it seems no less interesting to study the case which is opposite to one considered in the papers \cite{Vaeyrynen2017,Kurilovich2019,Kurilovich2019a,Pashinsky2020,Hsu2021,Probst2022}, namely, the case of a static magnetic defect with a large spin, which is robustly connected to the external environment.
   
Such a defect ensures the existence of a magnetic field in a small region of the HES, i.e. it allows elastic backward scattering without  tunneling coupling between the HES and defect. 
Possible experimental realization  is, for example, a potential dielectric ferromagnetic point contact with high magnetic stiffness and with the magnetic moment, whose direction is determined  by uniaxial anisotropy and the demagnetization tensor of the ferromagnet.   The possibility of creating static magnetic contacts to the HES has also been discussed  \cite{Khomitsky2022}.

Importantly, backscattering can occur not only  due  to  magnetic defect but also due to the interaction  with a charged puddle in the bulk of the topological insulator \cite{Vaeyrynen2013} or 
due to a point-like non-magnetic scatterer, with account for  the electron-electron interaction \cite{Sablikov2021}.

Here, we assume  the presence of backscattering static  defect (BD)  at the edge of the TI, which is  
 described  by the scattering matrix of the most general type 
 \begin{equation}
    \hat{S}_\mathrm{BD}= 
    \left( \begin{array}{cc}
     \cos \theta e^{i \alpha} &  i \sin \theta ~e^{i\varphi}\\
    i \sin \theta ~e^{-i\varphi} & \cos \theta e^{-i \alpha}\\
    \end{array} \right)  \,.
\label{SM}
\end{equation}
For the case of magnetic defect, we 
 neglect the back influence of the HES on the parameters of $ S_\mathrm{BD}$.     The backward scattering rate, $R_\theta= \sin^2 \theta, $  is determined by the quantity $\theta$ while the phase $\varphi$ has the meaning of the backward scattering phase on the BD.

Here we derive analytical expressions   for  a single BD  placed in the upper shoulder.
We consider interferometer with the lengths of  the upper and lower  shoulders given by $L_1$ and $L_2,$ respectively.  The BD is placed at position $x_0$ such that     $0<x_0<L_1.$
Using expression for scattering matrix \eqref{SM}, one can easily find  transfer matrix of  the defect 
\begin{equation}
\hat T_\mathrm{BD}= \frac {e^{-i \alpha }}{\cos \theta}   \left( \begin{array}{cc}
  1 &  i \sin \theta e^{-i \xi} \\
      -i \sin \theta e^{i \xi} & 1 
  \end{array} 
   \right),
     \label{eq:W}
\end{equation}
 where $\xi= \varphi- 2 k x_0$ and $k=\varepsilon/v_{\rm F}$ is the electron wave vector.  One can show that the forward scattering phase $\alpha$   can be fully incorporated into the shift of  $\phi,$  so that we put  $\alpha=0$ below.

\section{Shot noise and conductance}
\subsection{General equations}
 In this Section, we calculate  $\mathcal F$ for setups shown in Fig.~\ref{fig:setup} assuming that  a fixed bias voltage $V$ is applied to leads.  (In this paper, we assume  that the external impedance is zero and therefore the voltage does not fluctuate. For a finite impedance both current and voltage fluctuate as discussed in detail in Ref.~\cite{Blanter2000}.)
 
 We will consider the most interesting and easily realized case:
\be 
  \Delta \ll  T \ll e V.  
\label{ineq}
\ee

 The current noise is related with fluctuations of the  electric current with respect  to its average value
$\delta\hat{I} (t)=\hat{I} (t)- \langle\hat{I} \rangle$.
 Here $\hat I$ is the current operator (an analytical expression for $\hat I$ is given in \cite{Jong1997, Blanter2000}).

The current correlation function associated with the noise is defined by:
    \begin{equation*}
     {\mathcal   S} \left(t-t^\prime\right)=\frac{1}{2} \langle\delta\hat{I}(t)\delta\hat{I}(t^\prime)+\delta\hat{I}(t^\prime)\delta\hat{I}(t) \rangle \,.
    \end{equation*}
The Fourier transform of ${ \mathcal S}$ gives an expression for the noise power: $S(\omega)= 2 \int_{-\infty} ^\infty dt\,  e^{i \omega t}  {S(t)}  $ [the factor $2$ in this expression is a matter of convention, see Eq.~(1) in   Ref.~\cite{Jong1997} and the comment after Eq. (49) in   Ref.~\cite{Blanter2000}].

Spin-dependent transport through the two-terminal device is fully characterized by the matrix of transmission amplitudes $\hat{t}= t_{\alpha\beta}$ (here $\alpha$ and $\beta$ are the spin indices associated with the outgoing and the incoming electrons, respectively)~\cite{Jong1997,Blanter2000}:
\aleq{
    \mathcal S (\omega=0) &= 2G_0 \\ & \times  \int_{\mu}^{\mu+e V} d\varepsilon \, {\rm Tr} \left[{}\hat{\mathcal{T}}\left(\varepsilon\right)\left(1- \hat{\mathcal{T}}\left(\varepsilon\right)\right)\right] \,, 
    \label{eq:S}
}
    where  $G_0 = e^2/h$ conductance quantum and      \be \hat{\mathcal{T}} (\varepsilon) = \hat{t} (\varepsilon) \,  \hat{t}^\dagger (\varepsilon). \label{eq:T}\ee 
The averaged current, $I=\langle \hat I \rangle,$ 
and the Fano factor
are given by 
    \begin{eqnarray} 
    & e I= G_0 \int_{\mu}^{\mu + eV} d\varepsilon\, {\rm Tr} \left[ \hat{\mathcal{T}}\left(\varepsilon\right)\right]\,,
  \label{eq:current}
   \\
       & \mathcal F= \frac{\mathcal S(\omega=0) }{2eI}=\frac{\int_{\mu}^{\mu + eV}  d\varepsilon \, \mathrm{Tr}  \left[\hat{\mathcal{T}} (1- \hat{\mathcal{T}})\right]}{\int_{\mu}^{\mu + eV}  d\varepsilon \, \mathrm{Tr} [\hat{\mathcal{T}}]} \,.
    \label{eq:fano}
    \end{eqnarray}
The transmission amplitudes $t_{\alpha \beta} (\varepsilon)$ varies  on an energy scale on the order of the level spacing.
We focus on the  case, when the conditions \eqref{ineq} are satisfied. Then for the $\mathcal F$ we have
\begin{equation} \label{eq:fano1}
    \mathcal F=\frac{\mathrm{Tr}  \langle  \hat{\mathcal{T}} (1- \hat{\mathcal{T}})  \rangle_{\varepsilon}} {\mathrm{Tr}  \langle \hat{\mathcal{T}}  \rangle_{\varepsilon}} \,,
    \end{equation}
where the averaging is taken over a  temperature window in the vicinity of the Fermi level
 in the limit $T \gg \Delta$. In what follows we assume the linearized form of the spectrum, with $\varepsilon =  v_{\rm F} k$, so that the  energy averaging is reduced  to calculation of the integral $\left \langle \cdots \right \rangle_{\varepsilon} = 
\Delta^{-1} \int_0^\Delta d \varepsilon \,(\cdots) 
 = \frac{L}{2\pi} \int_0^{2\pi/L} d k \,(\cdots)$.

As seen from \eqref{eq:current},  the conductance is proportional to the transmission coefficient averaged over the spin and the energy: 
\aleq{
{\mathcal T} = \mathrm{Tr}  \langle \hat{\mathcal{T}}  \rangle_{\varepsilon}/2
\label{eq:T1}.}
 Introducing also the average
\aleq{
{\mathcal T}_2 = \mathrm{Tr}  \langle \hat{\mathcal{T}} \hat{\mathcal{T}}   \rangle_{\varepsilon},
 \label{eq:T2}}
one can write  
the Fano factor in the following form:
\begin{equation}
\mathcal F = 1-  \frac{{\mathcal T_2} }{2 {\mathcal T}},
\label{Sup-Fano}
\end{equation}
    
In the next two subsections we will discuss calculations  that  allow one to find $\mathcal F,$ using Eq.~\eqref{Sup-Fano}. Technically, the key idea is to present the matrix of transmission amplitudes in a special form, which allows performing  the energy averaging  analytically. 

\subsection{Conventional AB interferometer}

We start with  calculation of   $\mathcal F$  for CI having two arms with the lengths $L/2 \pm a$ and two contacts (not necessarily identical) described by  scattering matrices with real amplitudes [see Eq.~   \eqref{eq:SCI}]. We first find   convenient expression for the matrix  of transmission amplitude (that consists of one element for spinless CI, also conductance quantum is $G_0 =  e^2/h$ in this case),  substitute it into      the formula \eqref{eq:fano1}, and average over the energy.   

\subsubsection{Method of calculation}
We use the method developed in Ref.~\cite{Shmakov2013} to sum amplitudes with different winding numbers. We   classify all amplitudes of electrons going out of the ring by their direction, either clockwise, $(+)$, or anticlockwise, $(-)$. 
  For CI with identical contacts, the  amplitudes describing  processes of passing the ring  in opposite directions without extra revolutions  are given by
\aleq{
 \left( \begin{array}{c}
\beta_0^+ \\
\beta_0^-
\end{array} \right)  = 
t_{\mathrm{in}} t_{\mathrm{out}}  \left( \begin{array}{c}
e^{i(k - 2\pi \phi / L) (L/2 + a)} \\
e^{i(k + 2\pi \phi / L) (L/2 - a)}
\end{array} \right)\,.
}
All other trajectories finishing at the right lead can be obtained recursively, 
multiplying by a matrix describing processes which increase number of revolutions, clockwise or counterclockwise, by unity (for a  more detailed discussion, see Ref.~\cite{Shmakov2013}):
\aleq{
&\left( \begin{array}{c}
\beta_{n+1}^+ \\
\beta_{n+1}^-
\end{array} \right) = 
\hat{A}  \left( \begin{array}{c}
\beta_{n}^+ \\
\beta_{n}^-
\end{array} \right), \\
& \hat{A} = 
e^{ikL} \\
&\times \left( \begin{array}{cc}
t^2 e^{-i2\pi \phi} + t_b^2 e^{i2ka} & t t_b ( e^{-i2\pi \phi} + e^{i2ka}) \\
t t_b ( e^{i2\pi \phi} + e^{-i2ka}) & t^2 e^{i2\pi \phi} + t_b^2 e^{-i2ka}
\end{array} \right). 
}
The full transmission amplitude across the ring is given by:
\aleq{\label{eq:CIt}
{t}^{\rm CI}(\varepsilon) =  \sum_{n=0}^\infty \alpha \hat{A}^n \beta_0 = \alpha (1-\hat{A})^{-1} \beta_0\,, \\ 
\alpha = \left( \begin{array}{c}
1 \\
1
\end{array} \right).
}
Having particular expressions for $t^{\rm CI}(\varepsilon),$ we get
 ${\cal T}^\mathrm{CI}(\varepsilon) =|t^{\rm CI}(\varepsilon)|^2$ and  ${\cal T}_2^\mathrm{CI}(\varepsilon)=|t^{\rm CI}(\varepsilon)|^4$. The averaging over $\varepsilon$ is reduced to integration over the unit circle  in complex plane of $z = e^{ikL}$ and is easily performed by residues, see~\ref{app:aver}. 
 Most interesting and  physically transparent limiting cases are  presented  below.
In the most   general case of interferometer  with  
contacts  of different strength,  the obtained analytical expressions are  rather cumbersome   and we   present them   in the  \ref{app:CI}.

\subsubsection{Results (symmetric setup, identical contacts)}

 \begin{figure}
    \centering
    \includegraphics[width=0.8\linewidth]{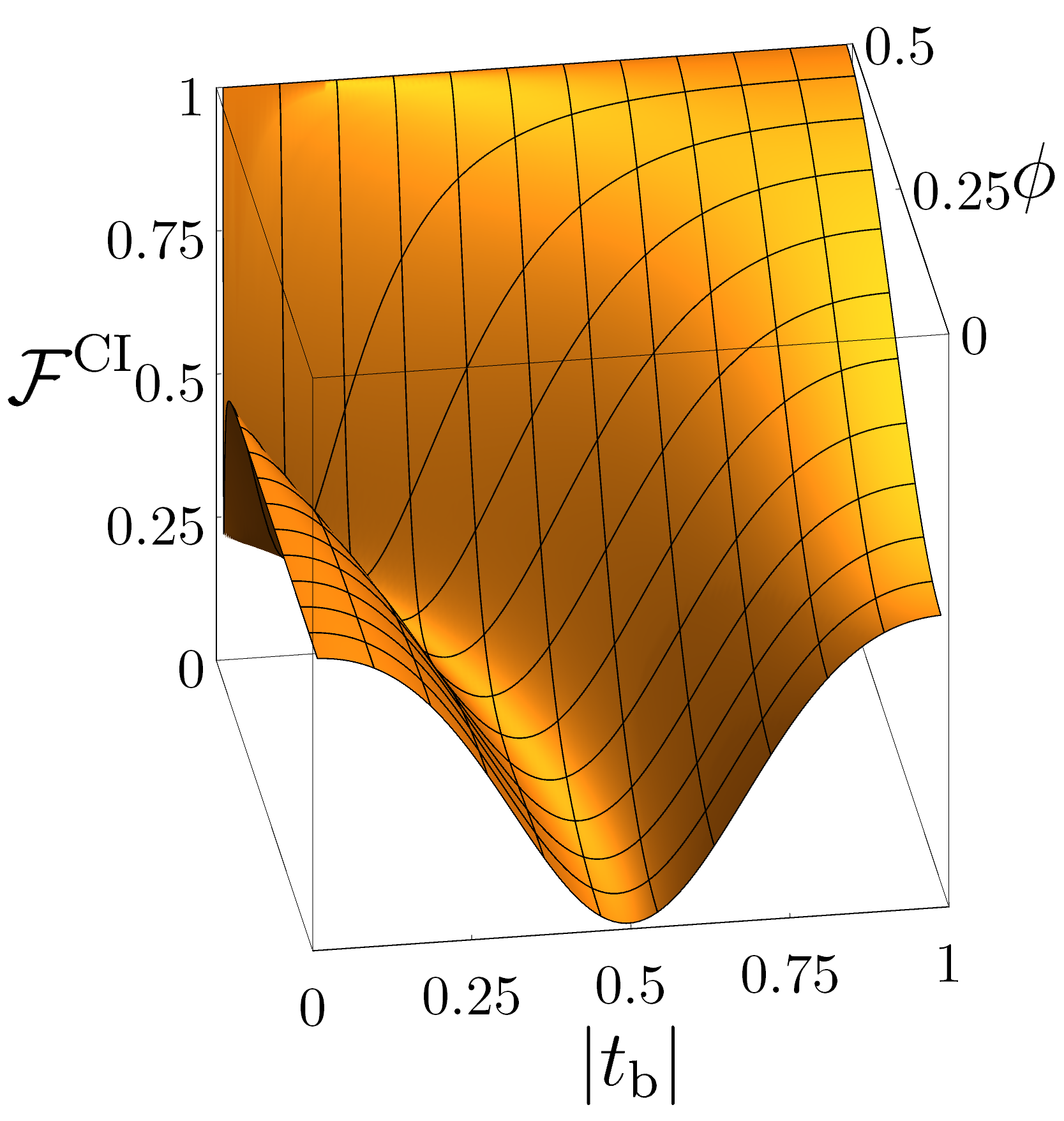}
\caption{The dependence of Fano factor, $\mathcal{F}^\mathrm{CI}(\phi,t_\mathrm{b})$, 
 on the  amplitude of backscattering on the contact, $t_\mathrm{b}$, and the magnetic flux, $\phi$, as described by Eqs.\ \eqref{eq:SCI}, \eqref{TCIsym}, \eqref{FFCIsym}. 
 }
\label{fig:fanoCI_3D}
\end{figure} 

 \begin{figure*}[ht]
    \centering
    \includegraphics[width=0.8\linewidth]{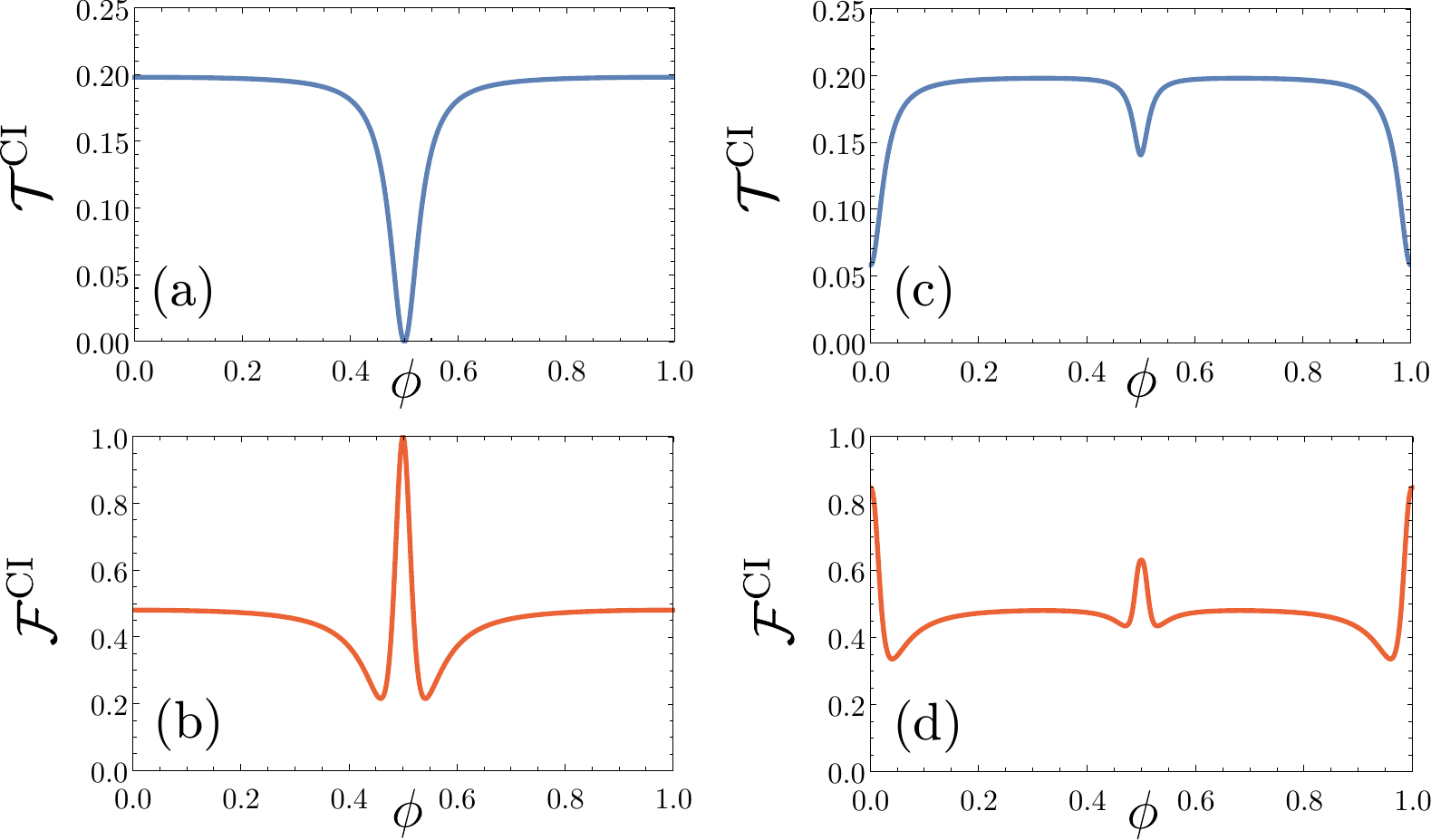}
\caption{The transmission coefficient (blue curves)  and the Fano factor (red curves) for  conventional    interferometer with two identical contacts.  In  {\it symmetric} interferometer with $a=0$ and weak tunneling coupling ($\gamma = 0.1$)   there is antiresonance in $\mathcal T^{\rm CI}$ and resonance in $\mathcal F^{\rm CI}$   at half-integer values of the flux [panels (a) and (b)].     In {\it asymmetric}  interferometer with $a\neq 0$ and weak tunneling coupling  ($\gamma = 0.1$, $k_F a = 1.0$), there also appear peaks at   integer values of the flux  [panels (c) and (d)]. 
With increasing $a,$  peaks    at $\phi=1/2+n$ decrease in amplitude and  become  narrower,  while peaks at $\phi=n$  grow and become wider (here, n is integer). 
Shape of the curve $\mathcal T^{\rm CI} (\phi)$  strongly depends on the asymmetry of the contacts. } 
\label{fig:fanoconv}
\end{figure*} 

For the case of a \emph{symmetric} interferometer with equal arm lengths ($a=0$)  and  identical contacts, the energy-averaged transmission coefficient  \cite{Dmitriev2010} is given by 
\aleq{
 {\mathcal{T}}^{\mathrm{CI}}(\phi,\gamma) =  \frac{2 \gamma  \cos ^2(\pi  \phi )}{\gamma ^2+\cos ^2(\pi  \phi )}\,.
 \label{TCIsym}
 }
Using formulas from 
\ref{app:aver} we find the Fano factor in the form (this equation was presented in Ref.~\cite{Niyazov2024} without derivation)
\aleq{
&\mathcal{F}^\mathrm{CI}(\phi,\gamma)  \!  =  
1\! \\
&-\! \frac{\cos^2(\pi \phi)\!\left[5 \gamma^2 \!+\!   \gamma^4\! +\! (1\!+\!\gamma^2) \cos^2(\pi \phi)\right]}{2\left[ \gamma^2 + \cos^2( \pi \phi) \right]^2}.
\label{FFCIsym}
}
Equation \eqref{FFCIsym} is illustrated 
in Fig. \ref{fig:fanoCI_3D}, where 
the dependence of Fano factor on the magnetic flux and on the  backscattering on the contact, $|t_\mathrm{b}|=\gamma/(1+\gamma).$   
For three types of tunneling contacts shown in  Fig.~\ref{fig:contact_conv}, Eq.~\eqref{FFCIsym} simplifies:
\aleq{
\mathcal{F}^\mathrm{CI}(\phi,0)=&\frac{1}{2} \,,\\
\mathcal{F}^\mathrm{CI}(\phi,\infty)=&1-\frac{1}{2} \cos ^2(\pi  \phi )  \,,\\
\mathcal{F}^\mathrm{CI}(\phi,1)=&\frac{4 \sin ^2(\pi  \phi )}{(\cos (2 \pi  \phi )+3)^2}\,.\\
\label{FanoCI}
}
The value $1/2$ for $\gamma \to 0$ is a standard value for the Fano factor of  tunneling level.  We  also note that the ``metallic''  contacts with $\gamma=1$, shows  both  minimal value  $\mathcal{F}^\mathrm{CI}=0$, at $\phi=0$,  and to the maximal value,  $\mathcal{F}^\mathrm{CI}=1$,  at $\phi \to 1/2.$ 

The dependencies of $ {\mathcal{T}}^{\mathrm{CI}}$ and $\mathcal{F}^\mathrm{CI}$ on the magnetic flux are shown in Fig.\ \ref{fig:fanoconv}a, \ref{fig:fanoconv}b.  The latter figure yields cross-section of Fig.~\ref{fig:fanoCI_3D} for fixed $\gamma.$  
In Fig.~\ref{fig:fanoconv}a we see that in the symmetrical interferometer, $a=0,$ the transmission coefficient and the Fano factor have, respectively,  sharp antiresonance and sharp resonance  at $\phi=1/2$  (and, hence, at any half-integer flux value).         
 \begin{figure}
    \centering
    \includegraphics[width=0.8\linewidth]{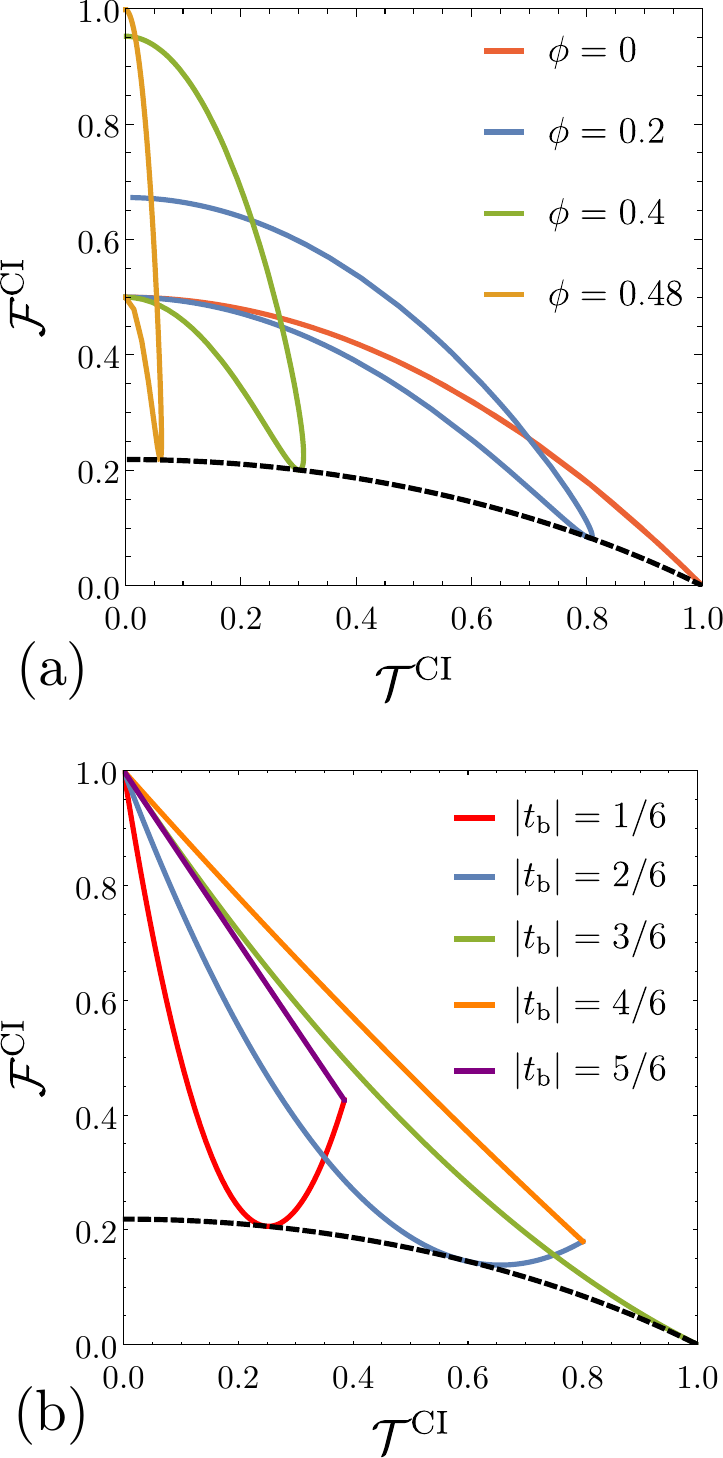}
\caption{Parametric plot $(\mathcal F^{\rm CI},  \mathcal T^{\rm CI}) $  for fixed $\phi$  and   $0<\gamma<\infty$ (a), for fixed $|t_{\rm b}|=\gamma/(1+\gamma)$ and $0<\phi<1$ (b).  
 }
\label{fig:fano_cond_CI}
\end{figure}

Several interesting properties of $ \mathcal{F}^\mathrm{CI}(\phi,\gamma)$ are to be mentioned here.   Firstly,
\be
\mathcal{F}^\mathrm{CI}(0,\gamma) = \mathcal{F}^\mathrm{CI}(0,1/\gamma)= \frac{1}{2}  
\left( \frac{1-\gamma^2}{1+\gamma^2}\right)  ^2 ,
\label{FFgammagamma}
\ee 
and also  $$\mathcal{T}^\mathrm{CI}(0,\gamma) = \mathcal{T}^\mathrm{CI}(0,1/\gamma)=\frac{2\gamma}{1+\gamma^2}.$$     It means that for zero flux, the  setup with the  contacts shown in     Fig.~\ref{fig:contact_conv}a with a  certain $\gamma$ is indistinguishable in terms of noise and conductance  from the setup with the  contacts shown in     Fig.~\ref{fig:contact_conv}b with   $\gamma_1=1/\gamma$. 

Let us discuss this point in more detail.  
As seen from Eqs.~\eqref{TCIsym} and \eqref{FFCIsym},  
 at  $\phi=0$ the Fano factor and the transmission coefficient       are related  as follows:
\aleq{
\mathcal{F}^\mathrm{CI}=\frac{1}{2} \left[1- (\mathcal{T}^{\mathrm{CI}})^2 \right] \,.
\label{F-T-CI}
}
This equation  is shown by red curve in Fig.~\ref{fig:fano_cond_CI} and is valid both for $\gamma <1$ (i.e. for contacts shown in Fig.~\ref{fig:contact_conv}a) and for $\gamma>1 $ (i.e. for contacts shown in Fig.~\ref{fig:contact_conv}b).  Specifically, changing $\gamma$ from $\gamma=0$ to  $\gamma=\infty,$ we pass the red curve twice: once  in the interval $0<\gamma<1$ and once again for     $1<\gamma<\infty.$ This is clearly seen from the cross-section of  Fig. \ref{fig:fanoCI_3D} at $\phi=0.$ Having in mind to compare below with the HI, we note that the role of the amplitude $t_\mathrm{b}$ is somewhat similar to the scattering on the magnetic impurity in case of HI, considered below. For the case of CI, dependence on $|t_{\rm b}|$ at $\phi=0$ (i.e. on the red curve  Fig.~\ref{fig:contact_conv}a) directly follows from Eq.~\eqref{FFgammagamma}: 
\[
\mathcal F^{\rm CI}_{\phi =0}= \frac{1}{2}
\frac{(|t_{\rm b}| -1/2)^2}{[(|t_{\rm b}| -1/2)^2+1/4]^2}
.\]
Importantly,  the  Fano factor has minimum with zero minimal value  in the ``metallic'' point,  $t_{\rm b}=1/2$ ($\gamma=1$). This property is lost in presence of magnetic flux, see Fig.~\ref{fig:fanoCI_3D}. 

For $\phi \neq 0,$ the red curve splits into two curves, as shown in Fig.~~\ref{fig:fano_cond_CI} by blue, green and orange lines.  Notice that the Fano factor can have two values for the same value of the transmission coefficient, corresponding to two different $\gamma.$   
Remarkably, there is a special  singular point, $\phi =1/2,$ where the conductance is exactly  zero (see \cite{Buettiker1984} and a more detailed discussion of this property in \cite{Dmitriev2010}). At this point,    the Fano factor becomes unity for fixed $\gamma$: 
\be 
\mathcal{T}^\mathrm{CI}(1/2,\gamma)=0, \quad  \mathcal{F}^\mathrm{CI}(1/2,\gamma)=1.
\label{CITFF1/2}
\ee 

The behavior of the Fano factor near this point is illustrated by the orange curve in Fig.~\ref{fig:fano_cond_CI}a, which corresponds to $\phi=0.48.$  
The transmission coefficient and the Fano factor  for $|\delta \phi| = |\phi -1/2| \ll 1,$  weak tunneling coupling,  $\gamma \ll 1, $  
and  arbitrary ratio
\[  x  = {\pi \delta \phi}/ { \gamma} .\]
are given by the following expressions
\begin{eqnarray}  
 \mathcal T^{\rm CI}(\delta \phi,\gamma) &   \approx 2 \gamma \frac{x^2}{1+x^2},
\\
\mathcal{F}^\mathrm{CI}(\delta \phi,\gamma) &\approx \frac{ 2-x^2 +x^4}{2 (1+x^2)^2}  \,,  
\label{F12CI}
\end{eqnarray}

 Hence, the value of $\mathcal{F}^\mathrm{CI}$ depends on the order of taking the limits $\delta \phi \to 0 $ and $\gamma \to 0,$ so that if we fix $\delta \phi $ and tend $\gamma$ to zero, then instead of Eq.\ \eqref{CITFF1/2} we get the usual expression for the tunnel contact: $$\mathcal{F}^\mathrm{CI} \to 1/2, \quad\gamma\to 0,\quad \delta\phi~ \mbox{ is fixed}  .$$

From Eqs.~\eqref{F12CI} we can get an analytical expression describing the orange line in Fig.~\ref{fig:fano_cond_CI}a. To this end, we  fix $\delta \phi$ and change tunneling coupling from $\gamma=0$ to relatively large value: $\delta \phi \ll \gamma \ll 1.$ Then, we get
at $ 0<\mathcal{T}^\mathrm{CI}<\pi|\delta \phi|$:
\be
\mathcal{F}^\mathrm{CI} \approx \frac{3}{4}- \frac{1}{2} \frac{\mathcal{T}^\mathrm{CI}}{\pi \delta \phi} \pm \frac{1}{4} \sqrt{1- \left( \frac{\mathcal{T}^\mathrm{CI}}{\pi \delta \phi}\right)^2} \,.
\ee
The sign in front of the last term distinguish two branches of the orange curve. The minimum of $\mathcal T^{\rm CI}$ is reached at the lower branch at $\mathcal T^{\rm CI}=15 \pi |\delta \phi|/16$ and is given by $\mathcal F^{\rm CI}=17/32,$ , which is the   
value of the dashed line in Fig.~\ref{fig:fano_cond_CI}a for $\mathcal{T}^\mathrm{CI}=0$.       

In Fig.~\ref{fig:fano_cond_CI}a we fixed $ \phi$ and changed $\gamma$ from zero to infinity. Fixing $\gamma$ instead and changing $ \phi$ in the interval  $0<\phi<1$, we  find the following expression connecting transmission coefficient and the Fano factor valid for any $\gamma$ in the  interval $0<\gamma <\infty$:
\begin{eqnarray} \label{FTgammafixed}
\mathcal{F}^\mathrm{CI} &=& 1- \frac{5+\gamma^2}{2} ~\frac{\mathcal{T}^\mathrm{CI}}{2\gamma} + 2 \left( \frac{\mathcal{T}^\mathrm{CI}}{2\gamma}\right)^2,
\\
&& 0<\mathcal{T}^\mathrm{CI}<\frac{2\gamma}{1+\gamma^2}.
\nonumber
\end{eqnarray}
This equation  has minimum at $\mathcal{T}^\mathrm{CI}_{\rm min}=(5 +\gamma^2) \gamma/4$ given by $\mathcal{F}^\mathrm{CI}_{\rm min}=[7-\gamma^2 (10+ \gamma^2)]/32.$ The family of the curves described by Eq.~\eqref{FTgammafixed} for different $\gamma$ is shown in Fig.~\ref{fig:fano_cond_CI}b. Note that the dashed curves in  panels  (a) and (b) of Fig.~\ref{fig:fano_cond_CI} coincide.
  From above discussion, we see that behavior of the noise of CI  is very singular for $\phi$ close to $1/2.$ This means that noise is very sensitive to the flux in the vicinity of this point.

\subsubsection{Results (asymmetric setup)}
The asymmetry of setup becomes important when $a \geq k_F^{-1}.$ For not too large voltages,     $eV   \ll v_F /a$, one can replace  $ka \rightarrow k_F a$ in  scattering amplitudes.    The expression for the transmission coefficient in asymmetric setup was obtained earlier in Ref.\ \cite{Dmitriev2010}. 
Analytical  expression for the  Fano factor in this case can be found by using formulas presented in \ref{app:aver} and \ref{app:CI}.  Its general form is too cumbersome and we present it here only for the flux values close to integer and half-integer number and for weak tunneling coupling.

For   $\delta \phi = |\phi - 1/2| \ll 1$ and $\gamma \ll 1$ the Fano factor reads:
\aleq{ 
\mathcal{F}^\mathrm{CI}_{1/2}(\delta \phi, \gamma) &= 
\frac{ 2-\bar{x}^2 + \bar{x}^4}{2 (1+\bar{x}^2)^2} \\
& + \frac{\sin^2(k_F a) [ 3\bar{x}^2 - \sin^2(k_F a) ]}{ 2(1 + 	\bar{x}^2)  [\bar{x}^2 + \sin^2(k_F a)]} ,  
\label{F1/2}
}
where   $$\bar{x}  = \frac{\pi\, \delta\phi}{\gamma \cos(k_F a)}.$$ 

Similar to  Eq.\ \eqref{F12CI}, the value of $\mathcal{F}^\mathrm{CI}_{1/2}$
at $\delta \phi\to 0$, $\gamma \to 0$ depends on the order of limits. Fixing  $\delta \phi$  and decreasing the   tunneling coupling to zero,  $\gamma \to 0 ,$ we get  $\bar x \to \infty,$ so that $\mathcal{F}^\mathrm{CI}_{1/2}\to 1/2$. 
Putting  $\delta \phi$ to zero first, we find
\be \mathcal{F}^\mathrm{CI}_{1/2} \to 1- \frac{\sin^2(k_{\rm F} a)}{2}.  \label{FCI12}\ee
It turns out, that the Eq.\ \eqref{F1/2} also describes the case of  small magnetic flux, $|\phi|\ll1$, if we 
set  $\delta \phi = \phi$ there and  replace $k_F a$ by $k_F a + \pi/2$.

As seen from Eqs.~\eqref{F1/2} and \eqref{FCI12}, the asymmetry of the device strongly affect the noise intensity  
even for fixed magnetic flux, so that changing $a$  within a small interval about the  Fermi wavelength, we change $\mathcal F^{\rm CI}$ by a factor on the order of two.

The dependencies of $ {\mathcal{T}}^{\mathrm{CI}}$ and $\mathcal{F}^\mathrm{CI}$ on the magnetic flux, $\phi$, are  shown in Fig.\ \ref{fig:fanoconv}c and \ref{fig:fanoconv}d.  Comparing these figures  with  Fig.~\ref{fig:fanoconv}a and \ref{fig:fanoconv}b plotted for symmetric case, we see that  new peaks at integer values of the  flux appear for $a \neq 0$ both in the transmission coefficient (antiresonances) and in the Fano factor (resonances) for integer values of flux. With increasing  asymmetry, the amplitudes of the peaks corresponding to half-integer values of the flux decrease, while the amplitudes of the peaks at integer flux values increase. We conclude that asymmetry of the device, which is always  present in  real experiments \cite{Shea2000,Piazza2000,Fuhrer2001,Keyser2003,Zou2007}, strongly modifies shot noise.
  
\subsubsection{Results (different tunneling couplings)}
In this subsection, we present results for  the case of symmetric setup, $a=0,$ but   different right and left contacts, characterized by parameter, $\gamma_\mathrm{R}$ and $\gamma_\mathrm{L}$, respectively. 
Using general equations presented in \ref{app:CI} and  performing standard energy averaging with the use of formulas from~\ref{app:aver}, one can  find the following expressions for the transmission coefficient and Fano factor: 
\begin{eqnarray*} 
 &{\mathcal{T}}^{\mathrm{CI}}(\phi,\gamma_\mathrm{R},\gamma_\mathrm{L}) =  \kappa^{-1/2} \frac{2 \bar \gamma  \cos ^2(\pi  \phi )}{ \bar \gamma ^2+\cos ^2(\pi  \phi )} \,,\\
&\mathcal{F}^\mathrm{CI}(\phi,\gamma_\mathrm{R},\gamma_\mathrm{L}) = 1-\kappa^{-1} \frac{ \cos ^2(\pi  \phi ) }{2 ( \bar \gamma^2+\cos^2( \pi  \phi ))^2} \\
&\times\left((1+ 4  \kappa)\bar \gamma^2+ \bar \gamma^4  +(1+\bar \gamma^2) \cos^2 (\pi  \phi )\right)
\,, 
\end{eqnarray*}
where $$
\bar \gamma  = \sqrt{\gamma_\mathrm{L} \gamma_\mathrm{R}}\,, \quad 
 \kappa = (\gamma_\mathrm{L} +  \gamma_\mathrm{R})^2 / 4 \gamma_\mathrm{L} \gamma_\mathrm{R}\,. $$

 The analog of  Eq. \eqref{F-T-CI} at  $\phi=0$ for different right and left contacts attains  the form   
\be 
\mathcal{F}^\mathrm{CI} 
= 1 - \frac{1}{2\kappa} - \frac {\kappa}{2}\left ({\mathcal{T}}^{\mathrm{CI}} \right)^2 \,.
\label{F-T-CI-2}
\ee 

\subsection{Helical AB interferometer}
 Next, we discuss the  Fano factor of the current noise in HI having two arms with lengths $L_1$ and $L_2.$ Just as in the case of CI we  start  formula for energy dependent amplitude $\hat t^{\rm HI}(\varepsilon),$ which is a matrix in spinful case, and then  perform energy averaging.  
The analytical expression  obtained for $\mathcal F^{\rm HI}$  is rather cumbersome, so we present it in the~\ref{app:helical} focusing in the main text on the most interesting limiting cases.

\subsubsection{Scattering matrix}
The matrix of the transmission amplitudes $\hat{t}^{\rm HI}$ from one contact to another is defined as follows
 \be 
 \left( \begin{array} {c}
 a^\uparrow  \\
 a^\downarrow 
\end{array} \right) = \hat{t}^{\rm HI}      \left( \begin{array} {c}
b^\uparrow  \\
 b^ \downarrow \end{array} \right)\,,
 \ee
where $(b^\uparrow, b^\downarrow)$ and $(a^\uparrow, a^\downarrow)$ are the amplitudes of incoming (from the right contact) and outgoing (to the left contact) waves, respectively (see Fig.~\ref{fig:setup}). This matrix has been obtained earlier~\cite{Niyazov2020}:
\aleq{ \label{eq:tmat}
\hat{t}^{\rm HI} &=  \frac{r^2 e^{2 \pi i \phi L_1/L}}{t^2} \left( \begin{array}{cc} e^{i k L_1} & 0 \\ 0&e^{-i k L_1} \end{array} \right) \\
&\times \left( \begin{array}{cc} t & 0 \\ 0&1 \end{array} \right)\hat g \left( \begin{array}{cc} t & 0 \\ 0&1  \end{array} \right)\,, \\
 \hat g &= \cos \theta \left[  \left( \begin{array}{cc} 0 & 0 \\ 0 & -1   \end{array} \right) \right. \\
 & \left.  + \frac{1}{2} \sum\limits_{\alpha=\pm1}   \frac{1+ \alpha \hat H}{1-t^2 e^{i( kL + \alpha  2\pi \phi_0)}}  \right]\,.
}
Here, $t$ and $r$ are scattering amplitudes entering matrix Eq.~\eqref{S-matrix},     $\phi_0$ is determined by the relation
\be
\cos(2\pi \phi_0)=\cos\theta \cos(2\pi \phi)\,,
\ee
and the matrix $\hat H$ is given by
 \be \hat H =
\left( \begin{array}{cc} a & b  e^{  i \xi}
\\ b e^{ - i \xi}  & -a \end{array} \right)\,,
\label{Hadamard}
\ee 
where $\xi$ is defined in Eq.~\eqref{eq:W}.
The coefficients
\begin{eqnarray}
&   
a= i\frac{
e^{-2 \pi i \phi} -\cos(2\pi \phi_0) \cos\theta}{\cos \theta
\sin(2 \pi \phi_0)} \,, 
\\
&
b= \frac{ e^{-2 \pi i \phi} \tan\theta}{ \sin (2\pi \phi_0) } \,,
\end{eqnarray}
 are related by the relation $a^2+b^2=1.$  

Calculations shows that asymmetry of the  arms does not play any role in the HI. Also, position  of MD $x_0$ drops out from final expressions for $\mathcal T^{\rm HI}$ and $\mathcal F^{\rm HI}.$
 We therefore  consider two cases: identical contacts (see next subsection and~\ref{app:HIid})   and non-identical contacts  (see~\ref{app:HInonid})  

Generalization of Eq.~\eqref{eq:tmat} for the case of different  tunneling contacts characterized by amplitudes $t_{\rm L}$ and $t_{\rm R}$ is straightforward.  One should make in this equation  the following  replacement:
$$t \to \sqrt{t_{\rm L} t_{\rm R}},\quad  r \to \sqrt{r_{\rm L} r_{\rm R}}, \quad \xi \to \xi + i \beta,$$ 
where   the parameter  describing the  difference of  tunneling contacts [see also \ref{app:HInonid}] is 
\aleq{ \label{eq:beta}
\beta= \frac{1}{2} \ln (t_\mathrm{R}/t_\mathrm{L}) \,.
}

\begin{figure}
    \includegraphics[width=0.8\linewidth]{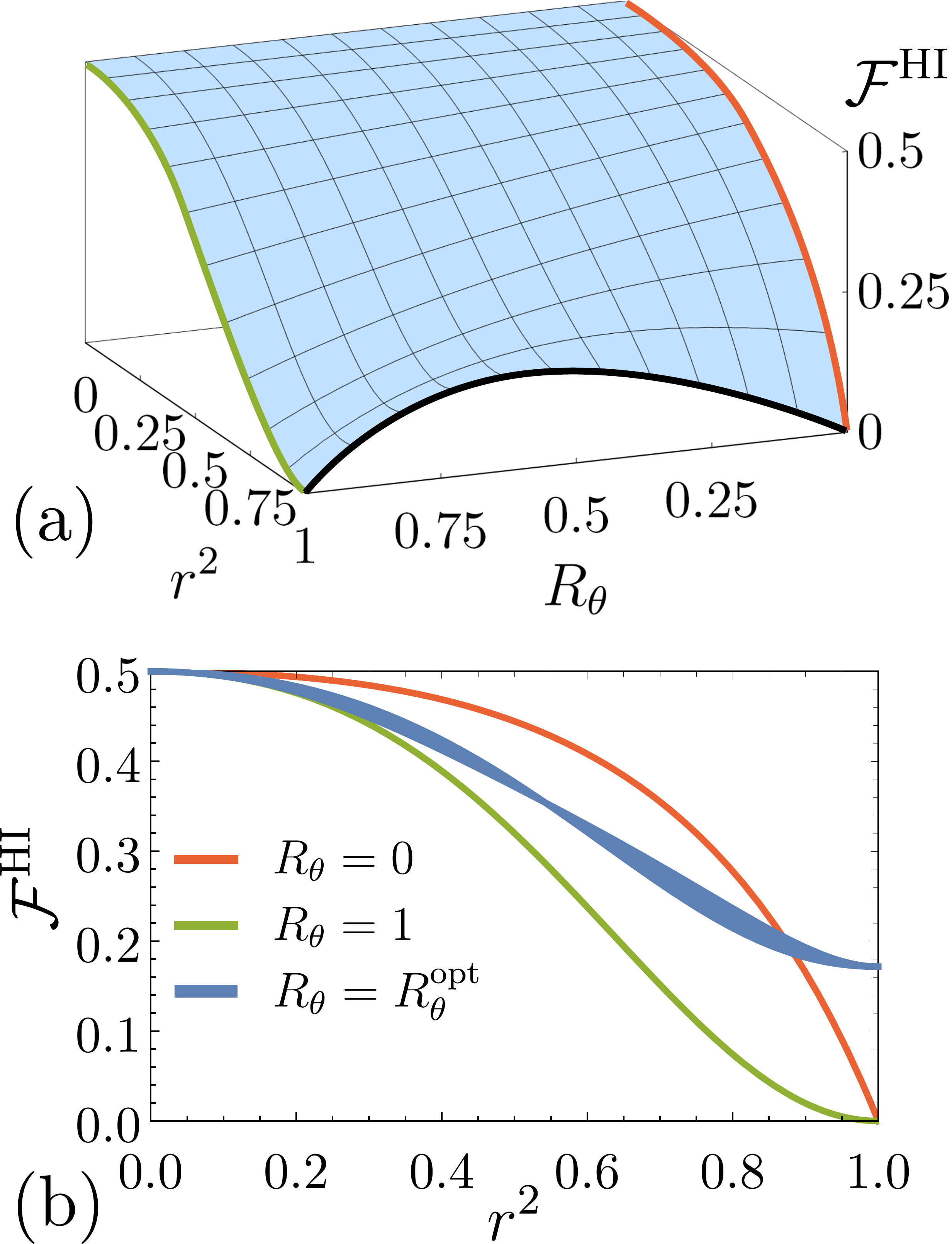}
    \centering
\caption{ 
Panel (a): The dependence of the Fano factor, $\mathcal{F}^\mathrm{HI} $, 
on $r^2$ and the strength of the magnetic impurity, $R_\theta$; for $r\to 1$ there an is optimal value, $R_\theta =R_\theta^{\rm opt},$ where the Fano factor has maximum. Panel (b): Cross-sections of panel (a) at different values of $R_\theta;$ dependence on the flux is weak and is illustrated for $R_\theta =R_\theta^{\rm opt}$  by broadening of the curve (shown by blue color) for flux belonging to interval $0<\phi<1$.   
  (these figures  were presented in Ref.~\cite{Niyazov2024} in  a slightly different way). The point $r=1 $ is ``noiseless''  both for ballistic case $R_{\theta}=0$ and the case of very strong defect $R_{\theta}=1.$   }   
\label{fig:FHI_r2}
\end{figure} 

\subsubsection{Results (identical contacts) }
The transmission coefficient  and the    Fano factor are obtained  with the use of Eqs.~\eqref{eq:T},  \eqref{eq:T1}, \eqref{eq:T2}, \eqref{Sup-Fano}, and \eqref{eq:tmat}.  
Direct energy averaging in Eqs.~\eqref{eq:T1} and \eqref{eq:T2}    yields \cite{Niyazov2021}
\aleq{
{\mathcal T}^{\mathrm{HI}} &=\tanh \lambda  \left[  1 \right.\\
& \left. -\frac{\sin ^2\theta  \sinh ^2\lambda  \cosh (2 \lambda )}{\cosh ^2(2 \lambda )-\cos ^2\theta  \cos ^2(2 \pi  \phi )} \right]
\label{eq:THI}
}
and \cite{Niyazov2024}
\begin{eqnarray}
 {\mathcal T}_2^{\mathrm{HI}} &= \tilde C \sum_{m,n=0}^3 \tilde A_{(m,n)} \frac{\sinh^{2 n}\lambda }{D^m} \,.
\label{T2-tilde}
\end{eqnarray}
Here
\aleq{
 D=&\sinh ^2(2 \lambda )+ R_\theta +(1 - R_\theta) R_\phi \,, \\
  \tilde C   =& \tanh \lambda / \cosh^2 \lambda, 
  \\
R_\theta=&  \sin^2 \theta, \quad 
R_\phi = \sin^2 (2 \pi \phi),
}
and the coefficients  $\tilde A_{(m,n)}$ depend on $R_\theta, R_\phi$  do not depend on $\lambda.$ Analytical  expressions for these coefficients  are presented in the \ref{app:HIid}. 
The  Fano factor is  connected with $\mathcal  T ^{\rm HI}$ and  $\mathcal  T ^{\rm HI}_2$  by  Eq.~\eqref{Sup-Fano}.

In Fig.~\ref{fig:FHI_r2}a, we present dependence of  the Fano factor,  
on  $r^2$ and the strength of the magnetic impurity, $R_\theta$ for $\phi=0$. For ``metallic'' contact, $r\to 1,$  dependence on the backscattering strength $R_\theta$ is non-monotonous.  Comparing this figure with analogous Fig.~\ref{fig:fanoCI_3D} for CI, we see that in both cases the Fano factor has   extremum  as a function  of  squared backscattering  amplitude. However,  for CI the Fano factor has minimum, while for HI  it has maximum. 

In Fig.~\ref{fig:FHI_r2}b, we plot dependence of $\mathcal F^{\rm HI} $ on $r^2$ for fixed $R_\theta.$  Two limiting dependencies, corresponding to absence of impurity, $R_\theta=0,$  and    very strong impurity, $R_\theta=1,$ are shown by orange and green line, respectively.  In both cases,  there is no dependence on the magnetic flux. For any other $R_\theta$ in the interval $0<R_\theta<1,$ the Fano factor depends on $\phi$ but this dependence is much weaker as compared to CI. Each curve corresponding to certain $R_\theta$ broadens as illustrated by blue curve corresponding to $R_\theta=R_\theta^{\rm opt}.$     

\begin{figure*}[ht]
    \centering
 \includegraphics[width=.8\linewidth]{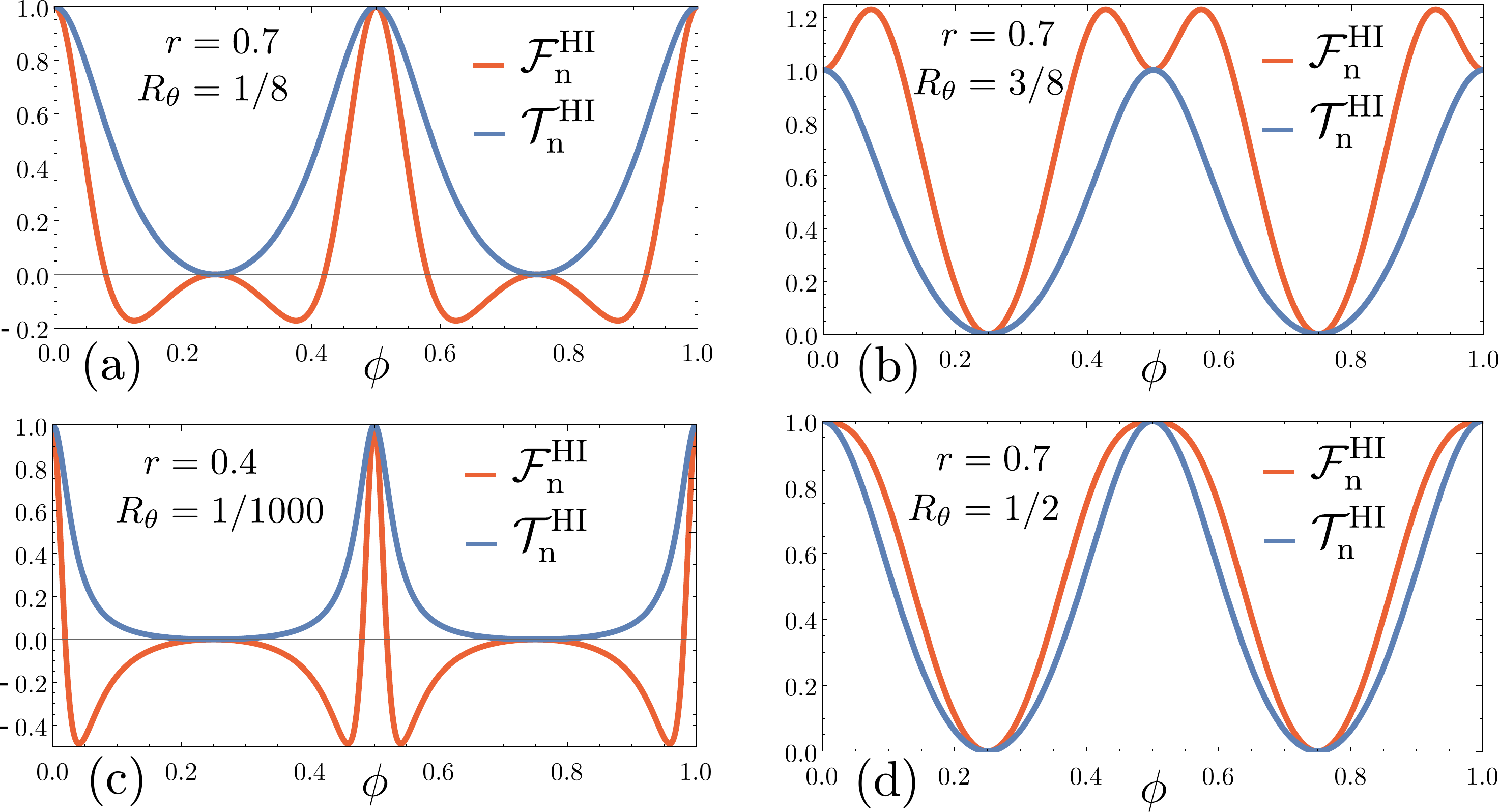}
\caption{Dependence of the normalized factor Fano, $\mathcal{F}_n^\mathrm{HI}$, and the normalized conductance, $\mathcal{T}_n^\mathrm{HI}$, for the helical interferometer on magnetic flux, $\phi$, at different strengths of scattering by a magnetic defect $R_\theta$ and tunneling amplitude, $r$. (a) $r=0.7, R_\theta=1/8$ (b) $r=0.7, R_\theta=3/8$ (c) $r=0.4, R_\theta=1/1000$ (d) $r=0.7, R_\theta=1/2$. 
Shape of the curve $\mathcal F^{\rm HI} (\phi)$  strongly depends on the  transparency of the contacts. }  
\label{fig:fano}
\end{figure*}

In order to emphasize this dependence one can consider normalized value 
\be \mathcal F_ {\rm n}^\mathrm{HI} (\phi) = \frac{\mathcal F ^\mathrm{HI}(\phi) - \mathcal F ^\mathrm{HI}(1/4)}{ \mathcal F ^\mathrm{HI}(0) - \mathcal F^\mathrm{HI} (1/4)}.
\label{normalized}
\ee 
The dependence of the normalized quantity $\mathcal F_ {\rm n}^\mathrm{HI}$ on the magnetic flux for several sets of parameters $R_\theta$ and $\lambda$  is shown in Fig.~\ref{fig:fano} along with the flux dependence of the normalized transmission coefficient. One can see sharp resonances both in  $\mathcal F_ {\rm n}^\mathrm{HI}$ and  $\mathcal T_ {\rm n}^\mathrm{HI}$  at weak tunneling coupling and small $R_\theta$ (see Fig.~\ref{fig:fano}c) which evolve into harmonic Aharonov-Bohm oscillations with increasing $\lambda$  and $R_\theta$    (see Fig.~\ref{fig:fano}d)
     
One can also plot dependence of the Fano factor  on the transmission coefficient similar to Fig.~\ref{fig:fano_cond_CI} for CI. Analogous dependence for  HI is shown in  Fig.~\ref{fig:fano_cond_HI}. Interesting difference of two pictures is that CI has a single noiseless point, $\mathcal F^{\rm CI}=0,$ corresponding to ideal transmission, $\mathcal T^{\rm CI}=1,$ while HI in addition to the analogous point with $\mathcal F^{\rm HI}=0, ~\mathcal T^{\rm HI}=1$   has one more noiseless point, corresponding to  $R_\theta=1,~\lambda=\infty.$ At this point $\mathcal F^{\rm HI}=0, ~\mathcal T^{\rm  HI}=1/2. $

Fig.~\ref{fig:fano_cond_HI} is one of the key results  concerning HI. Indeed, measuring   simultaneously   both conductance and noise and  using  Fig.~\ref{fig:fano_cond_HI}), one can find  the strength of the backscattering defect.   It is worth stressing that measuring the conductance only is insufficient to find the strength of the defect, because the conductance depends also on the properties of the contacts, which are actually not known to a sufficient precision. Importantly, by using both conductance and noise measurements, together with  Fig.~\ref{fig:fano_cond_HI} we can find $R_\theta$ without having detailed information about amplitudes $t$ and $r,$ characterizing the contacts.          It can help to identify   backscattering  mechanism  that apparently always exists in the realistic  experimental structures \cite{Olshanetsky2023a}.

\subsubsection{Limiting cases}
Here, we discuss several limiting cases allowing for simple  analytical description.  

\begin{figure}[h]
    \centering
    \includegraphics[width=0.7\linewidth]{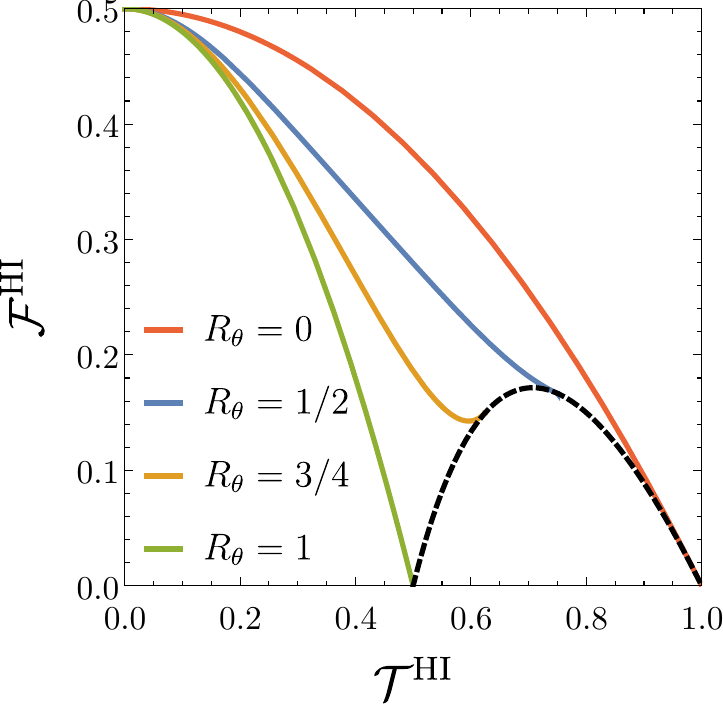}
\caption{ 
The combined dependence of Fano factor, $\mathcal{F}^\mathrm{HI} $, 
and the transmission coefficient,  $\mathcal{T}^\mathrm{HI} $, on the transparency of the contact, $\lambda$, for different strength of the magnetic impurity, $R_\theta$, and the magnetic flux $\phi =0.4$, according to Eqs.\  \eqref{eq:THI}, \eqref{T2-tilde}. 
The black dashed line borders the area of maximum transmission coefficient for given $R_\theta$. 
 }
\label{fig:fano_cond_HI}
\end{figure}

In the absence of a magnetic defect, $R_\theta=0$, interference effects are absent as we discussed in the introduction, so that expressions  for  the Fano factor and  transmission coefficient dramatically  simplify: 
\be
\mathcal T ^{\rm HI}=\tanh[\lambda], \quad \mathcal F^{\rm HI}= \frac{1}{2 \cosh^2[\lambda]}.
\ee
These equations not depend on $\phi$ and    are related by the same equation as for ballistic CI [compare with Eq.~\eqref{F-T-CI}]
\be
\label{eq:FanoCond}
\mathcal F^{\mathrm{HI}} =  \frac12 \left[ { 1-({\mathcal{T}}^{\mathrm{HI}})^2 }\right] .
\ee
This dependence is shown in Fig.~\ref{fig:fano_cond_HI} by red line

The formulas for the conductance and  Fano factor in  case of different  tunneling contacts can also be obtained in closed form and are  presented in \ref{app:HInonid}. 
  An analog of  Eq.\ \eqref{eq:FanoCond} in case of different  tunneling contacts, in contrast to CI case, cf. Eq.\ \eqref{F-T-CI-2} looks a bit  more cumbersome
\aleq{
 \mathcal F^{\rm HI}= \left[1- ({\mathcal{T}}^{\rm HI})^2 \right] \sqrt{\sinh^2(2\beta) +({\mathcal{T}}^{\rm HI})^2} \\
 / \left(\sqrt{\sinh^2(2\beta) +({\mathcal{T}}^{\rm HI})^2} + {\mathcal{T}}^{\rm HI} \cosh(2 \beta) \right) \,, 
}
with the asymmetry parameter $\beta = \frac{1}{2} \ln (t_\mathrm{R}/t_\mathrm{L}) $, Eq. \eqref{eq:beta}.       

The Fano factor  and transmission coefficient  also do not depend on the magnetic flux 
for the strong magnetic impurity, $R_\theta=1$ (green curves on the Fig. \ref{fig:FHI_r2}, accordingly). In this  case, we have
\begin{eqnarray}
&\mathcal T^{\rm HI}=\frac12 {\tanh[2 \lambda]},\quad \mathcal F^{\rm HI}=\frac{1}{2\cosh^2[2 \lambda]},
\\
& \mathcal F^{\rm HI}=\frac{1}{2} - 2 \left(  \mathcal T^{\rm HI} \right)^2.
\end{eqnarray}
Last dependence is shown by green line in Fig.~\ref{fig:fano_cond_HI}.

The flux dependence is also absent for the open interferometer, $r=1 ~(\lambda=\infty)$. In this case, we get 
\aleq{
&\mathcal{T}^\mathrm{HI}=1- \frac{R_\theta}{2},
\\
&\mathcal{F}^\mathrm{HI}=R_\theta\frac{1-R_\theta}{2-R_\theta}=\frac{ (2 \mathcal{T}^\mathrm{HI}-1) (1- \mathcal{T}^\mathrm{HI}) }{\mathcal{T}^\mathrm{HI}} \,.
\label{F-T}
}
Hence, the Fano factor has a maximum as a function of backscattering probability as is shown by the black curve in the Fig. \ref{fig:FHI_r2} (b). 

Equation \eqref{F-T}  has a simple physical meaning.  For $r=1,$ the windings are absent, and there are two parallel conductors, i.e. lower and upper arms of the HI.  The lower  arm, with spin $\downarrow$, is an ideal conductor with the transmission coefficient $\mathcal T_{\downarrow}=1$ while the  transmission through the  upper arm is controlled by the magnetic defect, $\mathcal T_{\uparrow} =\cos^2\theta, $ varying from  $0$ to $1.$ Then, we get
\be
\mathcal F^{\rm HI} =\frac{\mathcal T_{\uparrow} (1-\mathcal T_{\uparrow} ) }{1+\mathcal T_{\uparrow}}.
\label{Tuparrow}
\ee
Taking into account that the total transmission coefficient in this case is given by $\mathcal{T}^\mathrm{HI}=(\mathcal T_{\uparrow}+1)/2,$ we arrive to  Eq.~\eqref{F-T}.  We  have maximum $\mathcal F_{ \mathrm{max}}^{\rm HI}= 3-2\sqrt{2}\approx 0.17$
 at   $\mathcal T_{\uparrow,\mathrm{ max}}=\sqrt{2}-1$    and, respectively,   $R_\theta^\mathrm{max}=2-\sqrt{2} \approx 0.59$. The total transmission coefficient at this point is given by $\mathcal{T}^\mathrm{HI}_{\rm max}=1/\sqrt 2.$  We  notice that Eq.~\eqref{F-T} describes both black line in Fig.~\ref{fig:FHI_r2}a and dashed black line in Fig.~\ref{fig:fano_cond_HI}.  

 In order, to combine  here analytical formulas for  all relevant cases, we also present at the end of this section several equations obtained previously in Ref.~\cite{Niyazov2024}. For almost metallic contact, $t\ll 1$ ($\lambda \to \infty$), and 
for an arbitrary scattering strength on a magnetic defect,   $\mathcal F^{\rm HI}$ shows  weak oscillations   with $\phi$: 
\aleq{
\mathcal{F}^{\rm HI} \approx  \frac{R_\theta(1-R_\theta)}{2-R_\theta} + 2t^2
\\
+2t^4 \frac{ R_\theta (1-R_\theta) \left(10-12 R_\theta+3 R_\theta^2\right) }{(2-R_\theta)^2}\cos (4 \pi  \phi ).
}

More interesting is the opposite case of an almost tunnel contact: $t\to 1, r = \sqrt{1-t^2}\ll \theta.$ In this case,
\aleq{
\mathcal{F}^{\rm HI} \approx \frac{1}{2}-\frac{r^4}{8}\left[ 1+\frac{ 3R_\theta}{R_\phi + R_\theta  -  R_\theta R_\phi }\right]  .
\label{eq:t-to-1}}
This formula shows weak but sharp resonances with magnetic flux. These resonances are well pronounced   in the normalized Fano factor. 
Indeed, from  Eqs.~ \eqref{normalized} and \eqref{eq:t-to-1} for $\theta \ll 1$ we obtain
$$ \mathcal F_{\rm n} ^\mathrm{HI}\approx \frac{\theta^2}{\theta^2 + \sin^2(2 \pi \phi)}, $$ i.e. sharp resonances at $\phi=0$ and $\phi=1/2.$

\section{Conclusion}

In the previous sections we analyzed two types of  quantum interferometers based on conventional materials and helical edge states. We have demonstrated that  the shot noise   in both systems is substantially modified by quantum interference effects. It is essential that the interference effects survive in the high temperature limit,   $T \gg \Delta.$ The appearance of temperature-robust interference effects stems from trajectories interfering at any energy. In case of  symmetric CI for any trajectory entering the  interferometer and moving clockwise there is another trajectory moving counterclockwise, and these trajectories interfere destructively at $\phi=1/2.$
Such destructive interference is exact, i.e. the transmission amplitude is identically zero at all energies, $t(\epsilon,\phi=1/2) \equiv 0$ \cite{Buettiker1984} (see also the discussion of the consequences of this identity in \cite{Dmitriev2010,Dmitriev2015b}). 
The peculiarity of interference effects in  HI lies in the fact that they appear only in presence of backscattering defects. In this case trajectories interfering at any energy appear as well, namely, trajectories returning to the defect after $n$ revolutions clockwise and counterclockwise are always interfering.  The relative phase of such trajectories is  $4\pi n \phi$, and as a result the Aharonov-Bohm oscillation period in HI equals to $\Delta \phi =1/2,$  in contrast to CI, where the oscillation period is equal to   $\Delta \phi =1.$

To conclude, in this paper we discuss and compare noise characteristics of two types of Aharonov-Bohm interferometers, based on conventional materials and on helical edges states of TI. We focus on the case of relatively high temperatures, which partially smear the interference effects, all the while leaving many remarkable quantum features intact. In particular,  we show that even in case of higher temperatures the noise Fano factor reveals a characteristic  periodicity in the magnetic field. We demonstrate different patterns of this periodicity with resonances and anti-resonances at integer and half-integer values of the flux (see Fig.~\ref{fig:fanoconv} for CI and Fig.~\ref{fig:fano} for HI).  

One of the most important conclusions following from our calculations is  that the interference effects are  much more  pronounced  for CI and, consequently,  the influence of the magnetic field on the Fano factor is much stronger as compared to HI.  The reason is that the  interference  comes into play for CI already in the ballistic case, when impurities in the arms of CI are absent, while the Aharonov-Bohm peaks in  HI are proportional to the backscattering probability.

We show that asymmetrical placements of contacts results in strong modification of noise Fano factor in CI in contrast to HI. In particular, as seen from Eqs.~\eqref{F1/2} and \eqref{FCI12}, the asymmetry of the CI strongly affects the noise intensity  
even for fixed magnetic flux, so that changing difference of shoulder's lengths, $a,$  within a small interval about the  Fermi wavelength, we change $\mathcal F_{\rm CI}$ by a factor on the order of unity.  We also find that  the characteristics of the contact between the CI and the leads strongly modify the noise in this case, even in the ballistic regime inside the CI.  In turn, the noise  in case of HI is   strongly impacted by the backscattering defects inside the HI ring.  

We also demonstrate  that
the simultaneous measurement of conductance and Fano factor in HI allows one to determine the strength  of backscattering defects  that violate topological protection in the realistic setups. 
As for CI, it  can show different noise for the same transmission coefficient (see, for example, Fig.~\ref{fig:fano_cond_CI}). Hence, measuring,  complementary to conductance, also the Fano factor of CI would   help to experimentally determine the type of the contacts to leads (see   Fig.~\ref{fig:contact_conv}a and  Fig.~\ref{fig:contact_conv}b)  and the  strength of  tunneling coupling.  
  
\section*{Data availability statement}
All data that support the findings of this study are included
within the article (and any supplementary files).

\section*{Acknowledgments}

The work was carried out with financial support from the Russian Science Foundation (grant No. 25-12-00212), https://rscf.ru/en/project/25-12-00212/.

\onecolumngrid

\appendix

\section{Energy averaging} \label{app:aver}

Throughout the paper we average observable quantities over a small temperature window in the vicinity of the Fermi level  in the limit $T \gg \Delta$. For the linearized form of the spectrum, with $\varepsilon =  v_F k$, the  energy averaging is reduced at high temperatures to calculating the integral $\left \langle \cdots \right \rangle_{\varepsilon} = 
\Delta^{-1} \int_0^\Delta d \varepsilon \,(\cdots) 
 = \frac{L}{2\pi} \int_0^{2\pi/L} d k \,(\cdots)$. The averaging over $kL$, is reduced to integration over the unit circle $z = e^{ikL}$ in complex plane and is easily performed by residues.  The formulas which were used    for calculations  
 are listed below:
\aleq{
\left \langle \prod^4_{j=1}\frac{1}{(1 -\tau_j e^{ \alpha i k L})^{n_j}}\right \rangle_{\varepsilon} &=1\, , \quad n_j=0,\,1  \,, \\
\left \langle \frac{1}{(1 -\tau_1 e^{ \alpha i k L})}\frac{1}{(1 -\tau_2 e^{ -\alpha i k L})} \right \rangle_{\varepsilon} &= \frac{1}{1-\tau_1\tau_2}\, , \\
\left \langle \frac{1}{(1 -\tau_1 e^{ \alpha i k L})}\frac{1}{(1 -\tau_2 e^{ -\alpha i k L})}\frac{1}{(1 -\tau_3 e^{ -\alpha i k L})} \right \rangle_{\varepsilon} &= \frac{1}{(1-\tau_1\tau_2)(1-\tau_1\tau_3)}\, , \\
\left \langle \frac{1}{(1 -\tau_1 e^{ \alpha i k L})}\frac{1}{(1 -\tau_2 e^{ -\alpha i k L})}\frac{1}{(1 -\tau_3 e^{ -\alpha i k L})}\frac{1}{(1 -\tau_4 e^{ -\alpha i k L})} \right \rangle_{\varepsilon} &= \frac{1}{(1-\tau_1\tau_2)(1-\tau_1\tau_3)(1-\tau_1\tau_4)}\, , \\
\left \langle \frac{1}{(1 -\tau_1 e^{ \alpha i k L})}\frac{1}{(1 -\tau_2 e^{ \alpha i k L})}\frac{1}{(1 -\tau_3 e^{ -\alpha i k L})}\frac{1}{(1 -\tau_4 e^{ -\alpha i k L}) }\right \rangle_{\varepsilon} & \\
=  \frac{1-\tau_1\tau_2\tau_3\tau_4}{(1-\tau_1\tau_3)(1-\tau_2\tau_3)(1-\tau_1\tau_4)(1-\tau_2\tau_4)}\, ,&\\
}
where $\alpha=\pm 1$ and all $|\tau_j|<1$.

\section{ Conventional interferometer  with non-equal arms and   non-identical contacts  \label{app:CI}}
In this Appendix, we present generalization  of the tunneling amplitude $t^{\rm CI} (\varepsilon)$  [see   Eq.~\eqref{eq:CIt}] for the case of  non-identical  contacts having different tunneling couplings. The symmetric case is obtained by putting 
$t_{\mathrm{in,L}} = t_{\mathrm{in,R}} = t_{\mathrm{in}}$, etc.\ in equations presented below. 
\aleq{
t^{\mathrm{CI}}(\varepsilon) =&  \sum_{n=0}^\infty \vec \alpha \hat{A}^n \vec \beta_0 \,, \quad
\vec \alpha =  
\left( \begin{array}{c}
1 \\
1
\end{array} \right)  \,, \quad
 \vec \beta =  
\left( \begin{array}{c}
\beta_0^+ \\
\beta_0^-
\end{array} \right)  =
t_{\mathrm{in,L}} t_{\mathrm{out,R}}  
\left( \begin{array}{c}
e^{i(k - 2\pi \phi / L) (L/2 + a)} \\
e^{i(k + 2\pi \phi / L) (L/2 - a)}
\end{array} \right)\,, \\
\hat{A} =& 
e^{ikL}  
\left( \begin{array}{cc}
t_\mathrm{L} t_\mathrm{R} e^{-i2\pi \phi} + t_\mathrm{b,L} t_\mathrm{b,R}e^{i2ka} &  t_\mathrm{L} t_\mathrm{b,R} e^{-i2\pi \phi} + t_\mathrm{b,L} t_\mathrm{R}e^{i2ka} \\
t_\mathrm{L} t_\mathrm{b,R}e^{i2\pi \phi} + t_\mathrm{b,L} t_\mathrm{R}e^{-i2ka} & t_\mathrm{L} t_\mathrm{R} e^{i2\pi \phi} + t_\mathrm{b,L} t_\mathrm{b,R} e^{-i2ka}
\end{array} \right) \,. \\
}
After some algebra one can rewrite  expression for  $t^{\rm CI}(\varepsilon)$ as follows: 
\aleq{t(\varepsilon)^{\rm CI}=&t_{\mathrm{in,L}} t_{\mathrm{out,R}} e^{-i 2\pi \phi a/L} \left( \frac{Z e^{i(kL/2-\pi \phi)}}{1-\tilde t^2 e^{i{kL-2\pi \phi}}} +\frac{Z^\ast|_{\mathrm{L}\leftrightarrow \mathrm{R}} e^{i(kL/2+\pi \phi)}}{1-\tilde t^{\ast2} e^{i{kL+2\pi \phi}}} \right)\,, \\
Z =& i \frac{Z_1 Z_2}{Z_3} e^{i\pi \phi}\,,\\
Z_1=&Z_4-(t_\mathrm{b,L}-t_\mathrm{L}) (t_\mathrm{b,R} \sin (2 a k)+t_\mathrm{R} \sin (2 \pi  \phi ))-i( t_\mathrm{b,R} t_\mathrm{L} \cos (2 a k)+ t_\mathrm{b,L} t_\mathrm{R} \cos (2 \pi  \phi )) \,,\\
Z_2=&Z_4+(t_\mathrm{b,R}-t_\mathrm{R}) (t_\mathrm{b,L} \sin (2 a k)+t_\mathrm{L} \sin (2 \pi  \phi ))-i (t_\mathrm{b,L} t_\mathrm{R} \cos (2 a k)+ t_\mathrm{b,R} t_\mathrm{L} \cos (2 \pi  \phi )) \,,\\
Z_3=&2 Z_4 \left(t_\mathrm{b,L} t_\mathrm{R} e^{-i (a k+\pi  \phi )}+t_\mathrm{b,R} t_\mathrm{L} e^{i (a k+\pi  \phi )}\right) \,,\\
Z_4^2=& \left(t_\mathrm{b,L}^2-t_\mathrm{L}^2\right) \left(t_\mathrm{b,R}^2-t_\mathrm{R}^2\right)-(t_\mathrm{b,L} t_\mathrm{b,R} \cos (2 a k)+t_\mathrm{L} t_\mathrm{R} \cos (2 \pi  \phi ))^2  \,, \\
\tilde t^2 = & i Z_4+t_\mathrm{b,L} t_\mathrm{b,R} \cos (2 a k)+t_\mathrm{L} t_\mathrm{R} \cos (2 \pi  \phi ) \,.
\label{eq:tE-conv}}
Assuming that      $eV   \ll v_F /a$, one can replace  $ka \rightarrow k_F a$ in  this equation. 
Now we see that the expressions $|t^{\rm CI}(\varepsilon)|^2$,  $|t^{\rm CI}(\varepsilon)|^4$ contain the complex quantity  $e^{i kL}$ which is regarded as a new variable.  Then, one can perform averaging of these expressions over energy by using formulas derived in~\ref{app:aver}.

\section{Helical case} \label{app:helical}
\subsection{Identical contacts}
\label{app:HIid}
General formulas for the conductance and the noise intensity are obtained by averaging  the expressions ~\eqref{eq:S}, \eqref{eq:T}, \eqref{eq:current} and \eqref{eq:fano} in the main text  over the energy, using expressions for the amplitude \eqref{eq:tmat}. This result was obtained early in Ref. \cite{Niyazov2024} in different form.  It is   convenient to write the resulting formulas, using  the parameters $\lambda,$ $R_\theta$ and $R_\phi$ instead of the parameters $t,$ $\theta$ and $\phi$, according to the following definitions:
\begin{eqnarray}
&t=e^{-\lambda},\quad r=\sqrt{1-e^{-2\lambda}},\quad  0<\lambda<\infty, 
\\
&R_\theta=  \sin^2 \theta,
\\
&R_\phi = \sin^2 (2 \pi \phi). 
\end{eqnarray}
Direct calculation of energy averages in formulas \eqref{eq:T1} and \eqref{eq:T2} yields

\begin{eqnarray}
&{\mathcal T}^{\mathrm{HI}} = C \sum_{m,n=0}^1 A_{(m,n)} \frac{\sinh^{2 n}\lambda }{D^m} \,, 
\label{T1tilde}
\\
& {\mathcal T}_2^{\mathrm{HI}} = \tilde C \sum_{m,n=0}^3 \tilde A_{(m,n)} \frac{\sinh^{2 n}\lambda }{D^m} \,,
\end{eqnarray}
with 
\aleq{
D &=\sinh ^2(2 \lambda ) + R_\theta +(1 - R_\theta)R_\phi \,, \\
C   &=\tanh \lambda \,,\\
A_{(0,0)} & =1 - R_\theta/2 \,, \\
A_{(0,1)} & =0 \,, \\
A_{(1,0)} & =R_\theta(R_\theta+(1-R_\theta)R_\phi) / 2 \,, \\
A_{(1,1)} & =R_\theta \,,
\label{ABCD}
}
which leads to Eq. \eqref{eq:THI} and 
\aleq{
\tilde C   &= \tanh \lambda / \cosh^2 \lambda \,, \\
\tilde A_{(0,0)} &= 1+R_\theta(1-R_\theta/2) \,, \\
\tilde A_{(0,1)} &= 2-R_\theta(2-R_\theta)\,, \\
\tilde A_{(1,0)} &= R_\theta^2(R_\theta+R_\phi(1-R_\theta))/4\,, \\
\tilde A_{(1,1)} &= R_\theta(R_\theta-4(1-R_\theta)(3-2R_\theta)(1-R_\phi))/2\,, \\
\tilde A_{(2,0)} &= -R_\theta(2-R_\theta)(R_\theta+R_\phi(1-R_\theta))^2/2\,, \\
\tilde A_{(2,1)} &=-R_\theta (R_\theta(1-R_\phi)+R_\phi) \\
& \times (R_\theta(13-9R_\theta(1-R_\phi)-17R_\phi)+8R_\phi)/2\,, \\
\tilde A_{(2,2)} &= -R_\theta(R_\theta(9-7R_\theta(1-R_\phi)-11R_\phi)+4R_\phi)\,, \\
\tilde A_{(3,0)} &= -R_\theta^2(R_\theta+R_\phi(1-R_\theta))^3/4 \,, \\
\tilde A_{(3,1)} &= -3 R_\theta^2 (R_\theta+R_\phi(1-R_\theta))^3/2\,, \\
\tilde A_{(3,2)} &= -3R_\theta^2(R_\theta+R_\phi(1-R_\theta))^2\,, \\
\tilde A_{(3,3)} &= -R_\theta^2(4-2R_\theta(1-R_\phi)-2R_\phi) \\
&\times (R_\theta(1-R_\phi)+R_\phi)\,, \\
\tilde A_{(0,2)} &= \tilde A_{(0,3)}=\tilde A_{(1,2)}=\tilde A_{(1,3)}=\tilde A_{(2,3)}=0 \,.
\label{ABCD2}
}

\subsection{Non-identical contacts}
\label{app:HInonid}
In general case the tunneling rates in different contacts of the interferometer are described by unequal tunneling amplitudes
\aleq{
t_{\mathrm{R,L}}&= t e^{\pm \beta} \,, \quad |\beta|\leq \lambda\\
t_\mathrm{R} t_\mathrm{L} &= e^{-2\lambda} \,, \\
t_\mathrm{R}/t_\mathrm{L}&=e^{2 \beta} \,.
\label{t-beta}
}

In this case we may still use the expressions \eqref{T1tilde}, \eqref{T2-tilde}, but the coefficients should be redefined, 
$A_{(m,n)} \to A_{(m,n)} ^{\mathrm{u}} $, $\tilde A_{(m,n)} \to \tilde A_{(m,n)} ^{\mathrm{u}} $, as follows

\aleq{
C^{\mathrm{u}}   =& C \left(1-\frac{\sinh^2 \beta}{\sinh^2 \lambda}\right) \,,\\
A_{(0,0)}^{\mathrm{u}} =&A_{(0,0)}\,, \\
A_{(0,1)}^{\mathrm{u}} =&0 \,, \\
A_{(1,0)}^{\mathrm{u}} =&A_{(1,0)} + R_\theta \sinh^2\beta \,, \\
A_{(1,1)}^{\mathrm{u}} =&A_{(1,1)} + 2 R_\theta \sinh^2\beta \,,\\
\tilde C ^{\mathrm{u}}  =& \tilde C \left(1-\frac{\sinh^2 \beta}{\sinh^2 \lambda}\right)^2   \,, \\
\tilde A_{(0,0)}^{\mathrm{u}} =& \tilde A_{(0,0)} + 2R_\theta(2-R_\theta)  \sinh^2\beta \,, \\
\tilde A_{(0,1)}^{\mathrm{u}} =& \tilde A_{(0,1)} \,, \\
\tilde A_{(1,0)}^{\mathrm{u}} =& \tilde A_{(1,0)} + R_\theta (12-R_\theta(17-9R_\theta) \\
 &-3(1-R_\theta)(4-3R_\theta)R_\phi+4R_\theta \sinh^2\beta )\sinh^2\beta /2 \,, \\
\tilde A_{(1,1)}^{\mathrm{u}} =& \tilde A_{(1,1)}+R_\theta^2 \sinh^2(2\beta) \,, \\
\tilde A_{(2,0)}^{\mathrm{u}} =& \tilde A_{(2,0)}+R_\theta(R_\theta(1-R_\phi)+R_\phi)  \\
&\times (R_\theta(13-7R_\theta(1-R_\phi)-11R_\phi) \\
&-4(2-R_\phi))\sinh^2\beta/2 \\
&+5R_\theta^2(1-R_\theta)(1-R_\phi) \sinh^4\beta \,, \\
\tilde A_{(2,1)}^{\mathrm{u}} =& \tilde A_{(2,1)} - 4 R_\theta(2-R_\theta)(R_\theta(1-R_\phi)+R_\phi)\sinh^2\beta \\
&+10R_\theta^2(1-R_\theta)(1-R_\phi) \sinh^4\beta \,, \\
\tilde A_{(2,2)}^{\mathrm{u}} =& \tilde A_{(2,2)} -2R_\theta(R_\theta(9-7R_\theta(1-R_\phi)-11R_\phi) \\
&+4R_\phi)\sinh^2\beta\,, \\
\tilde A_{(3,0)}^{\mathrm{u}} =& \tilde A_{(3,0)} - R_\theta^2 (R_\theta(1-R_\phi)+R_\phi) \\
&\times((R_\theta(1-R_\phi)+R_\phi) 
 (2-R_\theta(1-R_\phi)-R_\phi) \\
 &+(4-3R_\theta(1-R_\phi)-3R_\phi)\sinh^2\beta)\sinh^2\beta \,, \\
\tilde A_{(3,1)}^{\mathrm{u}} =& \tilde A_{(3,1)} -6 R_\theta^2 (R_\theta(1-R_\phi)+R_\phi)(R_\theta(1-R_\phi)+R_\phi \\
&+(2-R_\theta(1-R_\phi)-R_\phi)\sinh^2\beta)\sinh^2\beta\,, \\
\tilde A_{(3,2)}^{\mathrm{u}} =& \tilde A_{(3,2)} - 12 R_\theta^2 (R_\theta(1-R_\phi)+R_\phi) (R_\theta(1-R_\phi) \\
&+R_\phi+\sinh^2\beta) \sinh^2\beta \,, \\
\tilde A_{(3,3)}^{\mathrm{u}} =& \tilde A_{(3,3)} - 2R_\theta^2 (R_\theta(1-R_\phi)+R_\phi) \sinh^2(2\beta) \,, \\
\tilde A_{(0,2)}^{\mathrm{u}} =& \tilde A_{(0,3)}^{\mathrm{u}}=\tilde A_{(1,2)}^{\mathrm{u}}=\tilde A_{(1,3)}^{\mathrm{u}}=\tilde A_{(2,3)}^{\mathrm{u}}=0 \,.
\label{ABCD2u}
}

\twocolumngrid


%

\end{document}